\documentclass[11pt]{article}

\usepackage{latexsym,mathrsfs}
\usepackage{amsmath,amssymb} 
\usepackage{amsthm,enumerate,verbatim}
\usepackage{amsfonts}
\usepackage[T1]{fontenc}
\usepackage{textcomp}
\usepackage{graphicx}
\usepackage{appendix}
\usepackage{booktabs}    
\usepackage{multirow}    
\usepackage{algorithm}
\usepackage{algorithmic}
\usepackage{makecell}

\usepackage{url}
\usepackage{colortbl}
\usepackage{nicematrix}
\usepackage{tikz}
\usepackage{hyperref}
\usetikzlibrary{arrows,shapes,positioning,shadows,trees}

\usepackage{bm}
\usepackage{mathtools, nccmath}

\newtheorem{definition}{Definition}
\newtheorem*{definition*}{Definition}

\DeclareMathOperator{\vect}{vec}

\newcommand{\bigO}{\mathcal{O}}

\definecolor{brightpink}{rgb}{1.0, 0.0, 0.5}
\newcommand{\ngc}[1]{{\color{brightpink} (\textbf{NG:} #1)}}

\definecolor{burntsienna}{rgb}{0.91, 0.45, 0.32}

\newcommand{\revise}[1]{{{\color{black} #1}}} 

\title{Algorithms for Boolean Matrix Factorization \\ 
using Integer Programming and Heuristics} 

\date{}

\author{Christos Kolomvakis \qquad Thomas Bobille \qquad Arnaud Vandaele \qquad Nicolas Gillis\thanks{Email: \{christos.kolomvakis, arnaud.vandaele, nicolas.gillis\}@umons.ac.be. The authors acknowledge the support by the European Research
Council (ERC, consolidator grant no 101085607, eLinoR). Thomas Bobille, MSc student from ENS Lyon, contributed to this project  during a semester internship at the University of Mons.} \\ 
Department of Mathematics and Operational Research \\ 
Facult\'e Polytechnique, Universit\'e de Mons \\ 
Rue de Houdain 9, 7000 Mons, Belgium   
	}

\begin{document}

\maketitle

\begin{abstract}
 Boolean matrix factorization (BMF) approximates a given binary input matrix as the product of two smaller binary factors. Unlike binary matrix factorization based on standard arithmetic, BMF employs the Boolean OR and AND operations for the matrix product, which improves interpretability and reduces the approximation error. It is also used in role mining and computer vision. In this paper, we first propose algorithms for BMF that perform alternating optimization (AO) of the factor matrices, where each subproblem is solved via integer programming (IP). We then design different approaches to further enhance AO-based algorithms by selecting an optimal subset of rank-one factors from multiple runs. To address the scalability limits of IP-based methods, we introduce new greedy and local-search heuristics. We also construct a new C++ data structure for Boolean vectors and matrices that is significantly faster than existing ones and is of independent interest, allowing our heuristics to scale to large datasets. We illustrate the performance of all our proposed methods and compare them with the state of the art on various real datasets, both with and without missing data, including applications in topic modeling and imaging. The code to run all experiments is available from 

\url{https://gitlab.com/ckolomvakis/boolean-matrix-factorization-ip-and-heuristics}. 
\end{abstract}

\textbf{Keywords.}  
Boolean matrix factorization, 
matrix completion, 
integer programming, 
alternating optimization,
greedy algorithms.

\section{Introduction}
\label{sec:intro}
Low-rank matrix approximations (LRMAs) are popular methods in machine learning and have been successfully used in a wide variety of applications, such as document classification, community detection, hyperspectral unmixing, and recommender systems, among others; see, e.g.,~\cite{markovsky2012low, udell2016generalized, NMF_book}. 
LRMAs perform dimensionality reduction by approximating an input data matrix as the product of two factors of smaller dimensions. 
Depending on the application, different models can be considered. Examples include principal component analysis (PCA) and its variants, such as sparse~\cite{zou2006sparse} and robust~\cite{candes2011robust} PCA, as well as nonnegative matrix factorization (NMF)~\cite{lee1999learning}. 
If the input matrix has entries in $\{0,1\}$, it makes sense to constrain the factors to take values in $\{0,1\}$ as well, leading to binary matrix factorization (bMF) and Boolean matrix factorization (BMF)~\cite{miettinen2008discrete, zhang2007binary, miettinen2009matrix_Thesis}. 
bMF uses standard addition and multiplication; hence, the approximation may produce elements that are not in $\{0,1\}$.
This restricts its interpretability and produces approximations with higher errors.
These issues are alleviated by BMF, which utilizes the Boolean-OR and Boolean-AND operations in matrix multiplication. 

This paper is organized as follows. 
In Section~\ref{sec:sec2}, we formally define BMF and illustrate its benefits compared to bMF; in particular, its ability to mine overlapping communities.
In Section~\ref{sec:AO}, we describe our proposed alternating optimization (AO) algorithm for BMF, where the subproblems are quadratic integer programs (IPs).
We also provide two initialization strategies. 
In Section~\ref{sec:Comb}, we provide an IP formulation to optimally combine several BMF solutions. 
In Section~\ref{sec:greedy}, to handle larger datasets, we introduce greedy algorithms to tackle BMF. We start by showing how we solve the Boolean least squares problem using a greedy algorithm and local search, 
and then we present greedy algorithms to combine several solutions, analogously to Section~\ref{sec:Comb}. In Section~\ref{sec:datastruct}, we describe our dedicated C++ data structure to deal with Boolean matrix products more efficiently.  
In Section~\ref{sec:experiments}, we provide numerical experiments on a variety of real-world datasets, including topic modeling and imaging. 

This work is an extended version of our conference paper \cite{KolomvakisMLSP}, whose content roughly corresponds to Section~\ref{sec:AO}. This paper provides the following new material:
\begin{itemize}
    \item A longer discussion on BMF: More specifically, we illustrate the use of BMF and its advantages over bMF. We also discuss identifiability results, related works, and other closely related models.
    
    \item New IP-based algorithms that improve upon our previous results from \cite{KolomvakisMLSP}, which can also handle matrices with missing data. 
    
    \item We propose new greedy algorithms that allow for scalability while still retaining competitive results with the state of the art.

    \item We design a dedicated new C++ data structure for Boolean data. 
    
    \item We conduct new experiments on real datasets, with and without missing elements.
    
    \item We show an application of BMF in topic modeling.
    
\end{itemize}


\section{Boolean matrix factorization (BMF)}
\label{sec:sec2}

Let us first define the matrix Boolean product.
\begin{definition}[Boolean matrix product]
	Given two Boolean matrices, $\mathbf{W} \in \{0,1\}^{m \times r}$ and $\mathbf{H} \in \{0,1\}^{r \times n}$, their \textit{Boolean matrix product} is denoted $\mathbf{W} \circ \mathbf{H} \in \{0,1\}^{m \times n}$ and is defined elementwise for all $i,j$ as 
	\begin{equation}\label{Bool_Prod}
		(\mathbf{W} \circ \mathbf{H})_{ij} = \bigvee_{k = 1}^r \mathbf{W}_{ik} \wedge \mathbf{H}_{kj} = \bigvee_{k = 1}^r \mathbf{W}_{ik}\mathbf{H}_{kj},
	\end{equation}
	where $\vee$ is the logical OR operation 
 (that is, $0 \vee 0 = 0$, $1 \vee 0 = 0 \vee 1 = 1$, and $1 \vee 1 = 1$) and $\wedge$ is the logical AND operation (that is, $0 \wedge 0 = 0$, $1 \wedge 0 = 0 \wedge 1 = 0$, and $1 \wedge 1 = 1$). For scalars that are in $\{0,1\}$, the Boolean AND and the standard multiplication are equivalent operations. 
 Interestingly, for binary matrices, the Boolean matrix product can be expressed in terms of the standard matrix product via the relation $\mathbf{W} \circ \mathbf{H} = \min(1,\mathbf{WH})$, where the minimum is taken entrywise and $\mathbf{WH}$ is the standard matrix product of $\mathbf{W}$ and $\mathbf{H}$. 
 \end{definition}

\noindent 

Let us now define the BMF problem. To make it more general, we allow the input matrix, $\mathbf{X}$, to have missing entries.
The matrix $\mathbf{M}$ is used to model incomplete datasets, that is, datasets for which some values are unavailable. The matrix $\mathbf{M}$ always has the same dimensions as the input matrix $\mathbf{X}$. If an element $\mathbf{X}(i,j)$ is observed, then $\mathbf{M}(i,j) = 1$; otherwise, it is equal to 0. For complete datasets (where all entries are observed),   $\mathbf{M}$ is the all-one matrix. 
\begin{definition}[BMF] 
	Given a Boolean matrix $\mathbf{X} \in \{0,1\}^{m \times n}$, a mask $\mathbf{M} \in \{0,1\}^{m \times n}$, and a factorization rank $r$, BMF aims to find matrices $\mathbf{W} \in \{0,1\}^{m \times r}$ and $\mathbf{H} \in \{0,1\}^{r \times n}$ that solve 
   \revise{ 
 	\begin{equation}
	\min_{\mathbf{W} \in \{0,1\}^{m \times r}, \mathbf{H} \in \{0,1\}^{r \times n}}	\| \mathbf{M} \, \odot \, \left( \mathbf{X} - \mathbf{W} \circ \mathbf{H} \right) \|_F^2,  
		\label{BMF}
	\end{equation}	
 where $\odot$ denotes the Hadamard or elementwise product between two matrices of the same dimensions, and $\| \cdot \|_F^2$ is the squared Frobenius norm. 
 }
\end{definition}

BMF identifies subsets of rows and columns of $\mathbf{X}$ that are highly correlated, since the entries equal to one in each binary rank-one factor $\mathbf{W}(:,k)\mathbf{H}(k,:)$ correspond to a rectangular submatrix of $\mathbf{X}$ that should contain many entries equal to one.
The Boolean OR operation among the rank-one factors provides more flexibility to the $\mathbf{W}$ and $\mathbf{H}$ factors, since any ones of the input matrix can be approximated by multiple rank-one factors. As an example, in community detection applications, this allows BMF to detect overlapping communities.
By contrast, bMF requires the rank-one factors to have disjoint supports; otherwise, the bMF solution $\mathbf{WH}$ has entries larger than 2. 
Applications of BMF include role mining  \cite{Role_mining_BMF,RMining_problem}, bioinformatics \cite{liang2020bem, haddad2018identifying} and computer vision \cite{lazaro2016hierarchical}.

\subsection{An illustrative example}  

Let us provide an illustrative example of BMF in community detection and how it is able to detect overlapping communities. We consider a real binary dataset of 101 animals with 17 characteristics (for example, "hairy", "can be airborne", "aquatic") \cite{UCI,BoolMF_IP}. We consider $r = 3$. The factors for a submatrix of the input are as follows:  
\begin{equation*}
\medmath{
    \begin{NiceArray}{c|p{0.5cm}|p{0.5cm}|p{0.5cm}|p{0.5cm}|p{0.5cm}|}
    &\Block[draw=white]{1-1}{\text{\textcolor{blue}{hair}}} & \Block[draw=white]{1-1}{\text{\textcolor{blue}{feathers}}} & \Block[draw=white]{1-1}{\text{\textcolor{blue}{eggs}}} & \Block[draw=white]{1-1}{\text{\textcolor{blue}{aquatic}}} & \Block[draw=white]{1-1}{\text{\textcolor{blue}{milk}}} \\
        \Block[draw=white]{1-1}{\textcolor{red}{bass}} &\Block[draw=black]{1-1}{0} & \Block[draw=black]{1-1}{0}& \Block[draw=black]{1-1}{1} & \Block[draw=black]{1-1}{1}& \Block[draw=black]{1-1}{0}  \\
        \Block[draw=white]{1-1}{\textcolor{red}{bear}}&\Block[draw=black]{1-1}{1} & \Block[draw=black]{1-1}{0}& \Block[draw=black]{1-1}{0} & \Block[draw=black]{1-1}{0}& \Block[draw=black]{1-1}{1}  \\
        \Block[draw=white]{1-1}{\textcolor{red}{chicken}}&\Block[draw=black]{1-1}{0} & \Block[draw=black]{1-1}{1}& \Block[draw=black]{1-1}{1} & \Block[draw=black]{1-1}{0}& \Block[draw=black]{1-1}{0}  \\
        \Block[draw=white]{1-1}{\textcolor{red}{gorilla}}&\Block[draw=black]{1-1}{1} & \Block[draw=black]{1-1}{0}& \Block[draw=black]{1-1}{0} & \Block[draw=black]{1-1}{0}& \Block[draw=black]{1-1}{1}  \\
        \Block[draw=white]{1-1}{\textcolor{red}{ostrich}}&\Block[draw=black]{1-1}{0} & \Block[draw=black]{1-1}{1}& \Block[draw=black]{1-1}{1} & \Block[draw=black]{1-1}{0}& \Block[draw=black]{1-1}{0}  \\
       \Block[draw=white]{1-1}{\textcolor{red}{sea horse}} &\Block[draw=black]{1-1}{0} & \Block[draw=black]{1-1}{0}& \Block[draw=black]{1-1}{1} & \Block[draw=black]{1-1}{1}& \Block[draw=black]{1-1}{0}  \\
    \end{NiceArray}
  \;=\;
    \begin{NiceArray}{|c|c|c|}
       \Block[draw=white]{1-1}{ } & \Block[draw=white]{1-1}{ } & \Block[draw=white]{1-1}{ }\\
        \hline 
        \Block[draw=black]{1-1}{\cellcolor{blue!20}1}& \Block[draw=black]{1-1}{\cellcolor{orange!20}0}  & \Block[draw=black]{1-1}{\cellcolor{green!20}0} \\
        \hline
        \Block[draw=black]{1-1}{\cellcolor{blue!20}0}& \Block[draw=black]{1-1}{\cellcolor{orange!20}0} & \Block[draw=black]{1-1}{\cellcolor{green!20}1}\\
        \hline
        \Block[draw=black]{1-1}{\cellcolor{blue!20}0} & \Block[draw=black]{1-1}{\cellcolor{orange!20}1} & \Block[draw=black]{1-1}{\cellcolor{green!20}0}\\
        \hline
        \Block[draw=black]{1-1}{\cellcolor{blue!20}0}& \Block[draw=black]{1-1}{\cellcolor{orange!20}0} & \Block[draw=black]{1-1}{\cellcolor{green!20}1}\\
        \hline
        \Block[draw=black]{1-1}{\cellcolor{blue!20}0}& \Block[draw=black]{1-1}{\cellcolor{orange!20}1} & \Block[draw=black]{1-1}{\cellcolor{green!20}0}\\
        \hline
        \Block[draw=black]{1-1}{\cellcolor{blue!20}1}& \Block[draw=black]{1-1}{\cellcolor{orange!20}0} & \Block[draw=black]{1-1}{\cellcolor{green!20}0}\\
        \hline
    \end{NiceArray}
    \;\circ\;
    \begin{NiceArray}{|c|c|c|c|c|}
        \hline
        \cellcolor{blue!20}0&\cellcolor{blue!20}0 &\cellcolor{blue!20}1 &\cellcolor{blue!20}1 &\cellcolor{blue!20}0\\
        \hline
        \cellcolor{orange!20}0&\cellcolor{orange!20}1 &\cellcolor{orange!20}1 &\cellcolor{orange!20}0 &\cellcolor{orange!20}0\\
        \hline
        \cellcolor{green!20}1&\cellcolor{green!20}0 &\cellcolor{green!20}0 &\cellcolor{green!20}0 &\cellcolor{green!20}1\\
        \hline
    \end{NiceArray}
}
 \medskip
 \end{equation*}

\noindent Each column of $\mathbf{W}$ and each row of $\mathbf{H}$ correspond to communities; in other words, each rank one-factor $\mathbf{W}(:,k)\mathbf{H}(k,:)$ represents a community. The columns of $\mathbf{W}$ assign animals to the communities, while the rows of $\mathbf{H}$ assign the characteristics to these communities. 
In this example, we observe that the first community has the bass and the seahorse (first column of $\mathbf{W}$), and the characteristics "eggs" and "aquatic" (first row of $\mathbf{H}$).
Hence, the first community represents the class of aquatic animals. 
Similarly, the second community that includes the ostrich and chicken and the characteristics "can be airborne" and "has tail" corresponds to birds. The third community includes the gorilla and the bear, and its characteristics include "has hair" and "produces milk"; hence, it corresponds to mammals. 

\begin{figure}[H]
    \centering
    \begin{minipage}{0.48\textwidth}
        \centering
        \includegraphics[width=0.9\textwidth]{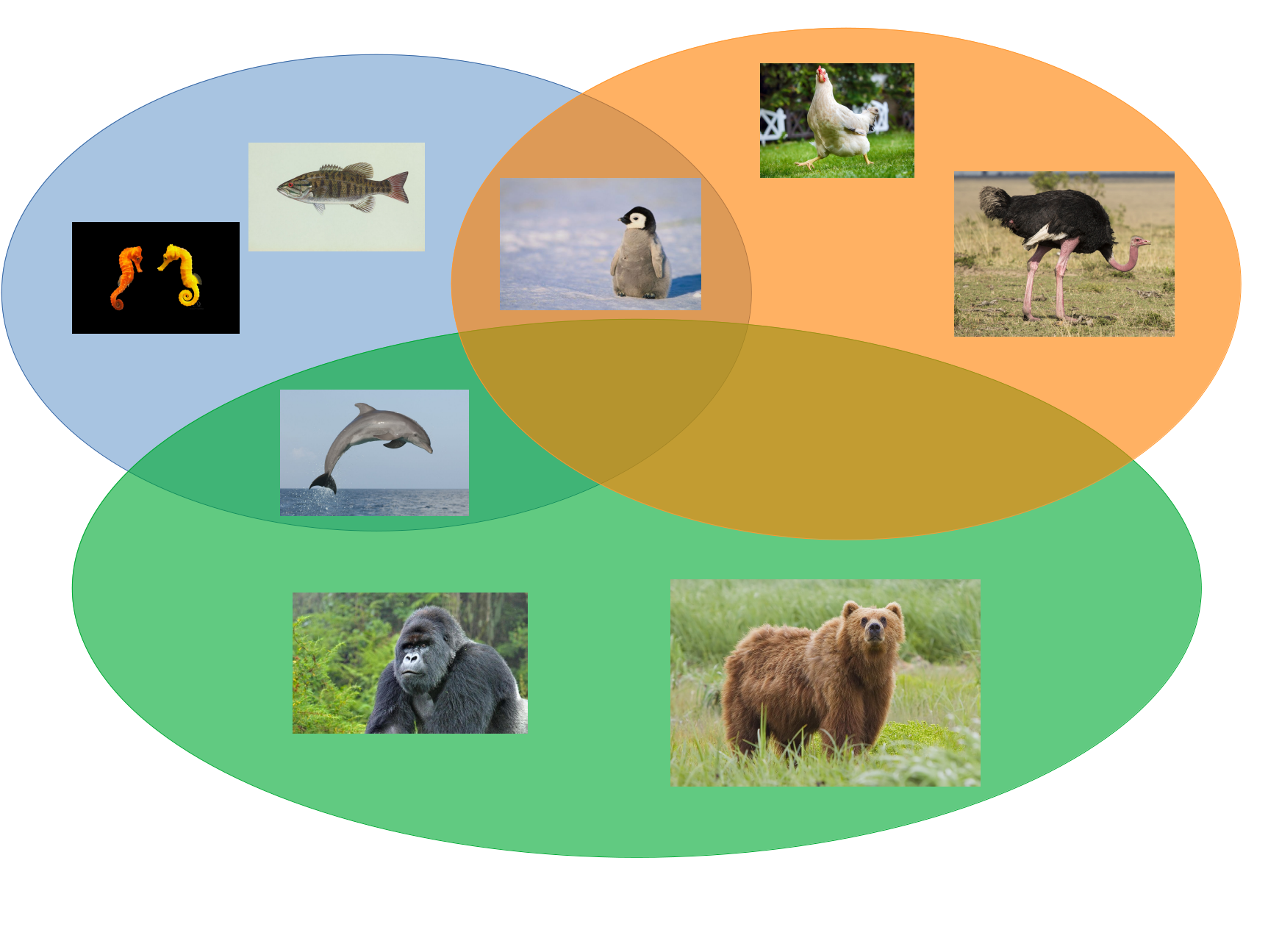} 
        \caption{Venn diagram showing the overlapping communities that animals are assigned to. }
        \label{FigVenn1}
    \end{minipage}\hfill
    \begin{minipage}{0.48\textwidth}
        \centering
        \includegraphics[width=0.9\textwidth]{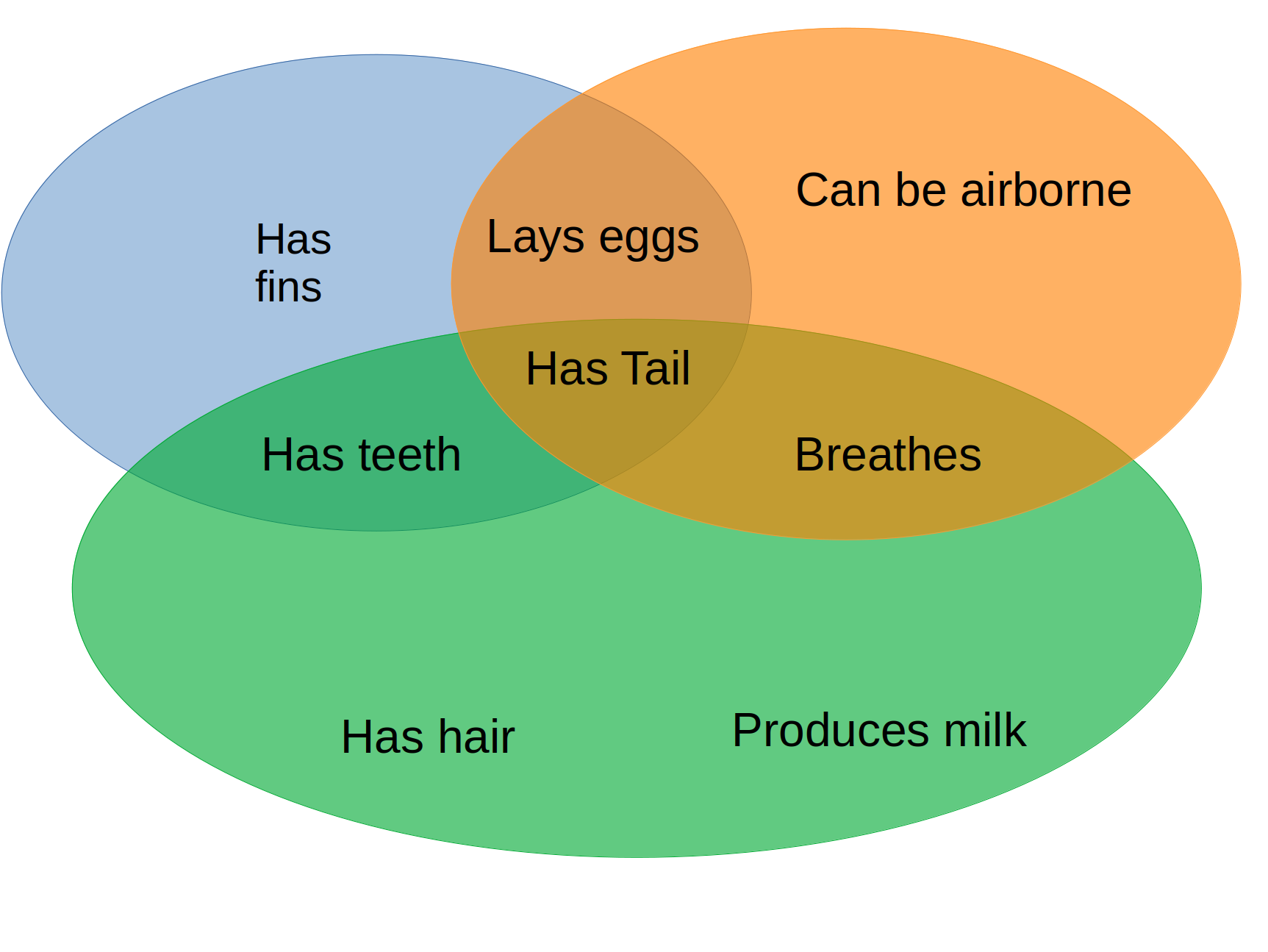} 
        \caption{Venn diagram showing the overlapping communities that animal characteristics are assigned to.}
        \label{FigVenn2}
    \end{minipage}
\end{figure}

In Figure \ref{FigVenn1}, we show a Venn diagram of the animals that are assigned to the communities. Note that some animals are assigned to multiple communities. 
Figure \ref{FigVenn2} represents a Venn diagram of the characteristics assigned to the communities. 
Note that not all characteristics need to be present for an animal assigned to a given community (this corresponds to a 0 entry in $\mathbf X$ being approximated by a 1). For example, 
the penguin cannot be airborne, but is still correctly classified as a bird.

\subsection{Previous BMF algorithms} 


 
In a similar manner to \cite{miettinen2020recent}, our brief review of past BMF algorithms puts them into three categories: (1) combinatorial approaches, (2)  continuous approaches where the BMF problem is relaxed in the continuous domain, 
and (3) miscellaneous approaches. 
 
 \paragraph{Continuous algorithms}
 Continuous approaches commonly (i) add penalties to incite the entries of $\bf W$ and $\bf H$ to belong to $\{0,1\}$, (ii) use a function on $\mathbf{\mathbf{WH}}$ (e.g., a threshold) to obtain a Boolean approximation, or (iii) combine both.
 
 The authors of \cite{miron2021boolean} consider a non-linear function for the Boolean matrix product and add penalty terms to enforce the factors to have binary elements. The authors then optimize on the nonnegative orthant. A similar work that optimizes in the nonnegative orthant and enforces the factors to be binary through a penalty term is in \cite{truong2021boolean}. Works using proximal gradient algorithms can be found in \cite{dalleiger2022efficiently,PRIMPING_Routine,c_salt}. Generalized versions of the Boolean matrix product are considered in \cite{generalized_miron,generalizedBMF_ICASSP}, where, instead of the Boolean OR, other logical operations are used, such as NAND and XOR. They then use a projected gradient to optimize the factors. \textit{FastStep} \cite{ribeiro-PAKDD2016} is a scalable algorithm that computes non-negative factors and then produces a Boolean approximation by applying a thresholding operator to the product $\mathbf{WH}$. The novelties of this algorithm lie in the computation of the cost function and the gradient, which reduce their costs and make it scalable.  
 
 Probabilistic continuous approaches to BMF also exist. In \cite{Bayesian_BoolMF}, a Bayesian approach to BMF is proposed; the authors fit the model using a metropolized Gibbs sampler. In \cite{pmlr-v48-ravanbakhsha16}, the BMF problem is recast as a maximum a posteriori (MAP) problem. The authors use probabilistic graphical models to implement a message passing algorithm that is scalable. Finally, the work in \cite{liang2019noisyincompletebooleanmatrix} proposes an expectation maximization (EM) algorithm that does not make any prior assumptions about the factors. The work also shows an application in breast cancer subtype classification. 

 \paragraph{Combinatorial algorithms} A seminal work on BMF is by Miettinen et al.~\cite{DBP_conf}, where the authors refer to the BMF problem as the \textit{Discrete Basis problem} (DBP). A greedy algorithm called ASSO is presented that computes the factors based on the correlations between the columns of the input matrix. It is also proven that approximating the BMF problem is NP-hard. An extended journal version of this work~\cite{miettinen2008discrete} presents improvements to ASSO and examines a closely related problem to DBP.

 Another class of algorithms makes use of \textit{formal concept analysis} \cite{FCA_I,FCA_II}, which is a mathematical framework used for data analysis. In \cite{BELOHLAVEK20103}, the authors present two algorithms: A greedy approximation algorithm based on the set covering problem. The second algorithm is a modified version of the first algorithm, in which the search space is significantly reduced. While in \cite{BELOHLAVEK20103} the algorithms are not named, in later works \cite{BELOHLAVEK201836,TRNECKA2022108895} they are referred to as GRECON and GRECOND, respectively.
 We also note that both of these algorithms are "from-below" algorithms. This means that if an entry in the input matrix is zero, then the corresponding entry of the approximation matrix will always be zero. More works that use formal concept analysis for BMF can be found in \cite{BELOHLAVEK201836,TRNECKA2022108895}.

 More recently, G{\"u}nl{\"u}k et al.~\cite{BoolMF_IP} relied on sophisticated integer programming relaxations of BMF, combined with dedicated optimization strategies to solve large instances (namely, column generation).

 \paragraph{Miscellaneous works} The papers~\cite{approx_schemes,A_PTAS,l0_approx,streamingPTAS,parametrised_theor_bmf} provide BMF algorithms with theoretical guarantees, including approximation algorithms. Finally, a recent work that uses collaborative neurodynamic optimization can be found in \cite{LI2022142}. \\
 
 We refer the interested reader to the recent detailed survey \cite{miettinen2020recent} on BMF.
 The authors not only analyze various algorithms for BMF, but also discuss problems that are equivalent to BMF (such as biclique covering \cite{bicliques} and the set cover problem \cite{tiling_databases,algos_book}), theoretical results, and several open problems.

\section{IP-based algorithms for BMF} \label{sec:AO}

Most algorithms for LRMAs rely on iterative block coordinate descent methods: the subproblem in $\mathbf{H}$ is solved for $\mathbf{W}$ fixed, and vice versa. The reason is that these subproblems are typically convex. For BMF, this is, of course, not the case. However, the advances in IP solvers, such as Gurobi~\cite{gurobi}, allow one to tackle medium-scale problems efficiently.

\subsection{IP formulation for BMF subproblems}

Assuming $\mathbf{W}$ is fixed in~\eqref{BMF}, we would like to solve the following Boolean least squares (BoolLS) problem in $\mathbf{H}$, that is, solve 
\[
\min_{\mathbf{H} \in \{0,1\}^{r \times n}} 
\|\mathbf{X} - \min(1, \mathbf{WH}) \|_F^2. 
\] 
Because of the nonlinearity in the objective, this cannot be solved directly with standard IP solvers. 
Note that the problem in each column of $\mathbf{H}$ is independent: 
\begin{equation} \label{BoolLS} 
\min_{\mathbf{H}(:,j) \in \{0,1\}^{r}} 
\| \mathbf{X}(:,j) - \min(1, \mathbf{WH}(:,j)) \|_2^2. 
\end{equation}
For simplicity, let $\mathbf{h} = \mathbf{H}(:,j)$ and \mbox{$\mathbf{x} = \mathbf{X}(:,j)$}. Given $\mathbf{W}$ and $\mathbf{x}$, we need to solve (\ref{BoolLS}). 
Introducing the variable $\mathbf{z} = \min(1, \mathbf{Wh})$, \eqref{BoolLS} can be reformulated as follows: 
\begin{equation} \label{bool_ls}
\min_{\mathbf{h} \in \{0,1\}^r, \mathbf{z} \in \{0,1\}^m}  \|\mathbf{x} - \mathbf{z}\|_2^2 
\quad 
\textrm{s.t.}  \quad 
\frac{\mathbf{Wh}}{r} \leq \mathbf{z} \leq \mathbf{Wh}. 
\end{equation} 
In fact, for $\mathbf{W}$, $\mathbf{h}$ and $\mathbf{z}$ binary, $\frac{\mathbf{Wh}}{r} \leq \mathbf{z} \leq \mathbf{Wh}$ if and only if $\mathbf{z} = \min(1,\mathbf{Wh})$, since  $\mathbf{Wh} \in \{0,1,\dots,r\}^m$. 
Now \eqref{bool_ls} is a convex quadratic optimization problem with linear constraints over binary variables. Such problems can be solved with commercial software, 
and  we make use of Gurobi~\cite{gurobi}. Other alternatives that are commercial IP solvers include CPLEX \cite{cplex2009v12} and Mosek \cite{mosek}. An open source alternative is GLPK \cite{Oki2012GLPKL}.
 For the case of missing data, we consider the following cost function:

\begin{equation} \label{bool_ls_missing}
\min_{\mathbf{h} \in \{0,1\}^r, \mathbf{z} \in \{0,1\}^m}  \|\mathbf{m} \odot (\mathbf{x} - \mathbf{z})\|_2^2 
\quad 
\textrm{s.t.}  \quad 
\frac{\mathbf{Wh}}{r} \leq \mathbf{z} \leq \mathbf{Wh}, 
\end{equation} 
where $\odot$ denotes the Hadamard (or elementwise) product, and $\mathbf{m}$ is a weight matrix such that $\mathbf{m}_i = 1$ if $\mathbf{x}_i$ is not missing, and 0 otherwise. 

\subsection{AO for BMF}

We can now solve BMF via AO over the factors $\mathbf{W}$ and $\mathbf{H}$ alternatively. Since $\|\mathbf{X} - \min(1,\mathbf{WH})\|_F^2
= \| \mathbf{X}^\top - \min(1,\mathbf{H}^\top \mathbf{W}^\top)\|_F^2$, the problem in $\mathbf{W}$ for $\mathbf{H}$ fixed has the same form. 
We update $\mathbf{H}$ in a column-by-column fashion by solving the independent BoolLS of the form~\eqref{BoolLS}, 
and similarly for $\mathbf{W}$ row by row. Algorithm~\ref{alg:bool_AO} summarizes the AO strategy. The algorithm is rewritten, in comparison to \cite{KolomvakisMLSP}, to also include the case of input with missing data. We have added a safety procedure within AO (steps~\ref{step5}-\ref{step9}): it may happen that some rows of $\mathbf{H}$ are set to zero (for example, if $\mathbf W$ is not well initialized). 
In that case, we reinitialize these rows as the rows of the residual $\mathbf R = \max\big(0,\mathbf X - \min(1, \mathbf{W}_{i-1} \mathbf{H}_i)\big)$ whose entries have the largest sum. This guarantees that the error will decrease after the update of $\mathbf{W}$.  


 



	

\begin{algorithm}[ht] 
\caption{AO algorithm for BMF - \texttt{AO-BMF}}
\begin{algorithmic}[1] \label{alg:bool_AO} 
\REQUIRE Input matrix $\mathbf{X} \in \{0,1\}^{m \times n}$, initial factor matrix $\mathbf{W}_0 \in \{0,1\}^{m \times r}$, mask matrix \\ $\mathbf{M} \in \{0,1\}^{m \times n}$, maximum number of iterations maxiter. 

\ENSURE $\mathbf{W} \in \{0,1\}^{m \times r}$ and $\mathbf{H} \in \{0,1\}^{r \times n}$ such that $\mathbf{M} \odot \mathbf{X} \approx \mathbf{M} \odot (\min(1,\mathbf{WH}))$. 
 
    \medskip  

\STATE $i = 1$, $e(0) = \|\mathbf{X}\|_F^2$, $e(1) = \|\mathbf{X}\|_F^2-1$. 

\WHILE{$e(i) < e(i-1)$ and $i \leq$ maxiter}
\STATE $\mathbf{H}_i = 
\texttt{BoolLS}(\mathbf{X},\mathbf{W}_{i-1},\mathbf{M})$. 
\STATE $\mathcal{K} = \{k \ | \ \mathbf{H}_i(k,:) = 0\}$. \label{step5}
\IF{ $\mathcal{K} \neq \emptyset$ }
    \STATE $\mathbf R = \max\big(0,\mathbf{M} \odot (\mathbf X - \min(1, \mathbf{W}_{i-1} \mathbf{H}_i))\big)$. 
    \STATE $\mathbf{H}_i(\mathcal{K},:) = \mathbf{R}(\mathcal{I},:)$, where $\mathcal{I}$ contains the indices 
    \STATE  of the $|\mathcal{K}|$ rows of $\mathbf{R}$ with largest sum. \label{step9} 
\ENDIF 
\STATE $\mathbf{W}_i = \texttt{BoolLS}(\mathbf{X}^\top, \mathbf{H}_{i}^\top,\mathbf{M}^\top)^\top$. 
\STATE $i = i + 1$. 
\STATE $e(i) = \| \mathbf{M} \odot (\mathbf{X} - \min(1, \mathbf{W}_i \mathbf{H}_i)) \|_F^2$. 
\ENDWHILE
\RETURN $(\mathbf{W},\mathbf{H}) = (\mathbf{W}_i, \mathbf{H}_i)$. 
\end{algorithmic}  
\end{algorithm}

\noindent Typically, AO needs a very small number of iterations to converge, given the combinatorial nature of the problem.
\vspace{-0.15mm}


\subsection{Initialization of AO}\label{sec:ao_init}

In this section, we provide two initialization strategies for AO-BMF, that is, Algorithm~\ref{alg:bool_AO}. \vspace{-3mm}
\paragraph*{Randomly selecting columns or rows of $\mathbf{X}$}

 AO-BMF only requires $\mathbf{W}$ to be initialized. By symmetry, it could also be initialized only with $\mathbf{H}$, starting the AO algorithm by optimizing over  $\mathbf{W}$. 
A simple, fast, and meaningful strategy to initialize AO-BMF is to initialize $\mathbf{W}$ (resp.\ $\mathbf{H}$) with a subset of the columns (resp.\ rows) of $\mathbf{X}$, that is, 
set  
$\mathbf{W} = \mathbf{X}(:,\mathcal{K})$ (resp.\ $\mathbf{H} =  \mathbf{X}(\mathcal{K},:)$)  where $\mathcal{K}$ is a randomly selected set of $r$ indices of the columns (resp.\ rows) of $\mathbf{X}$. 

\paragraph*{NMF-based initialization} 

The second initialization we propose relies on NMF. NMF approximates $\mathbf{X}$ with $\mathbf{WH}$ where $\mathbf{W}$ and $\mathbf{H}$ are nonnegative. 
We use an NMF algorithm from~\url{https://gitlab.com/ngillis/nmfbook/} which by default initializes the entries of 
 $\mathbf{W}$ and $\mathbf{H}$ with the uniform distribution on $[0,1]$. Once an NMF solution is computed, we binarize it using the following two steps: 
 \begin{itemize}
     \item Normalize the columns of $\mathbf{W}$ and the rows of $\mathbf{H}$ such that $\max(\mathbf{W}(:,k)) = \max(\mathbf{H}(k,:))$ for all $k$, using the scaling degree of freedom in NMF, that is,  $\mathbf{W}(:,k) \mathbf{H}(k,:) = (\alpha \mathbf{W}(:,k))(\alpha^{-1} \mathbf{H}(k,:))$ for $\alpha > 0$. 
     
     \item Set the entries of $\mathbf{W}$ and $\mathbf{H}$ to 0 or 1 using a given threshold $\delta$, which is generated uniformly at random  in the interval $[0.3, 0.7]$.  An alternative would be to use a grid search approach to determine $\delta$, similar to~\cite{zhang2007binary, truong2021boolean}. 
However, we observed that selecting $\delta$ at random performs better on average.  
 \end{itemize}   

\noindent It is important to note that if the input matrix has missing elements, we can only use the NMF-based initialization.

\subsection{Combining multiple BMF solutions}  \label{sec:Comb}

In this section, we present algorithms aimed at improving the performance of AO-BMF. With BMF being an NP-hard problem, it is expected that the solutions we find are local minima. To obtain better solutions, we use combining schemes, that is, we gather multiple solutions $(\mathbf{W},\mathbf{H})$ and try to pick the $r$ best rank-one factors to minimize the error. 

\subsubsection{MS-Comb-AO} \label{sec:MSCOMB}

Algorithm~\ref{alg:bool_AO}, AO-BMF, is able to generate locally optimal solutions for~(\ref{BMF}) relatively quickly, in the sense that they cannot be improved by optimizing $\textbf{W}$ or $\textbf{H}$ alone. 
A natural first approach to generating good solutions to BMF is to use multiple initializations and keep the best solution. 

However, it is possible to combine a set of solutions more effectively. 
Assume we have generated $p$ rank-$r$ BMFs: 
$\mathbf{W}_1 \mathbf{H}_1, \dots, \mathbf{W}_p \mathbf{H}_p$. This gives $rp$ binary rank-one factors, namely \mbox{$\mathbf{W}_\ell(:,k)\mathbf{H}_\ell(k,:)$} for $k=1,\dots,r$ and $\ell=1,\dots,p$.
Let us denote these rank-one binary matrices by $\mathbf{R}_i$ for $i=1,\dots,N$ with \footnote{In practice, we delete duplicated rank-one factors so that $N \ll rp$.} $N = rp$. 
To generate a better rank-$r$ BMF, we can pick $r$ rank-one binary factors among the $\mathbf{R}_i$'s by solving the following combinatorial problem: 
\[
\min_{\mathbf{y} \in \{0,1\}^N} \Big\|\mathbf{X} - \min\big( 1 , \sum_i y_i \mathbf{R}_i \big) \Big\|_F^2 \; \text{ such that } \; \sum_i y_i = r.  
\] 
The variable $y \in \{0,1\}^N$ encodes the $r$ selected rank-one factors, that is, $y_i = 1$ if $\mathbf{R}_i$ is selected in the BMF. 
As for BoolLS, we can reformulate this problem as a quadratic IP: 
        \begin{align}
               \label{Comb_BoolMF}
 \min_{\mathbf{y} \in \{0,1\}^N, 
 \mathbf{Z} \in \{0,1\}^{m\times n}} 
            &  \left\| \mathbf{X} - \mathbf{Z} \right\|_F^2 
 \; \; \text{ such that }  \; \;
 \sum_{i=1}^N y_i = r \; \text{ and } \; 
  \frac{\sum_{i=1}^N y_i \mathbf{R}_i}{r} \leq \mathbf{Z} \leq \sum_{i=1}^N y_i \mathbf{R}_i. 
        \end{align}
         
        Denoting $y_i^*$ the optimal solution of~\eqref{Comb_BoolMF},  
the rank-$r$ BMF obtained, $\sum_i y_i^* \mathbf{R}_i$, is guaranteed to be at least as good as all the solutions $\{\mathbf{W}_i \mathbf{H}_i\}_{i=1}^p$, since they are feasible solutions of~\eqref{Comb_BoolMF}. 
Once a solution combining several BMFs is computed, we further improve it using AO. 

We refer to this algorithm, namely generating $p$ solutions with AO-BMF, then combining them solving~\eqref{Comb_BoolMF}, and then applying AO to that solution, as MS-Comb-AO. \\
Finally, similarly to (\ref{bool_ls_missing}), we can also apply this combining scheme to the case where we have missing data by making a few adaptations to our original algorithm.




\subsubsection{Tree BMF}\label{subsec:hiertreebmf}

We now propose another method that uses MS-Comb-AO as a building block, which we refer to as Tree BMF. It is motivated by the fact that, when $N$ is large, solving~\eqref{Comb_BoolMF} can be computationally demanding. 
Hence instead of combining all solutions at once, we are going to build a tree and combine less solutions in each node of the tree. 

The method requires two parameters: the depth of the tree as well as the number of solutions for the MS-Comb-AO algorithm used by the leaf nodes. The procedure is as follows: At the leaf nodes, we solve multiple MS-Comb-AO according to the number of solutions specified by the user. Then, the leaf nodes send their best solutions to their parent, and the parent combines them using only two pairs of $(\mathbf{W}, \mathbf{H})$ factors. This continues until we reach the root node, where, after the combination step is performed, the final factors are returned. As a last step,  AO-BMF is used to improve the solution.

\section{Greedy and heuristic algorithms for BMF} \label{sec:greedy}

As we show in Section~\ref{sec:experiments}, the IP-based methods proposed in Sections \ref{sec:AO} and \ref{sec:Comb} are competitive compared to the state of the art. However, using Gurobi prevents us from scaling our inputs. For this reason, we now present novel greedy heuristics for the computation of BMFs. 
Greedy algorithms are often used as a means to tackle hard problems quickly. The trade-off is that, usually, the solutions are suboptimal \cite{algos_book}. We start this section by proposing greedy algorithms to solve the Boolean least squares problem, which we then use to solve BMF. We then propose combining schemes, similar to those proposed in  Section~\ref{sec:Comb}, that we can use to further improve our solutions. These combining algorithms will also not make use of an IP solver.  
Finally, we propose in Section~\ref{sec:datastruct} a new C++ data structure for Boolean vectors and matrices that is significantly faster than existing ones. 

\subsection{Greedy algorithm and local search for Boolean LS}

We start our discussion with the first step in designing heuristic algorithms for BMF, namely, greedy algorithms to solve Boolean LS combined with local search.

\subsubsection{Greedy algorithm}



    
\noindent Recall the BoolLS problem: 
\begin{equation} \label{BoolLS_2} 
\min_{\mathbf{h} \in \{0,1\}^{r}} 
\| \mathbf{x} - \mathbf{W \circ h} \|_2^2. 
\end{equation}
Our proposed greedy algorithm works as follows: $\mathbf{h}$ is initialized at the all-zero vector, that is, $\mathbf{h} = {\bf{0}}_r$. The algorithm sets entries to one in a greedy manner: 
We test setting each zero entry of $\mathbf{h}$ to one. 
The entry that reduces the error the most is set to 1 if it reduces the error of the previous solution; otherwise, the algorithm terminates and returns the current $\mathbf{h}$. Any entry set to 1 is not set to 0 later on. 

At each iteration of this algorithm, the selection of the best entry requires comparing the $r$ vectors of $\mathbf{W}$, in $\bigO(rm)$ operations, and updating the corresponding element of $\mathbf{h}$ in $\bigO(1)$ operations. 
The number of iterations is bounded by the number of elements of $\mathbf{h}$ because we set at least one component to 1 per iteration. Therefore, the total complexity is in $\bigO(mr^2)$ operations. 

For $r=1$, the greedy algorithm is optimal as the two solutions, 0 and 1, are explored. It turns out that, for $r=2$, it can happen that the greedy algorithm is not optimal, although this is not likely to happen in practice. 
Here is an example: 
\begin{align*}
 \bf{x} & = [0, 0, 0,  1,1, 1,1], \\
\bf{W}(:,1) & = [1, 1, 1, 1,1, 0,0], \\  
\bf{W}(:,2) & = [1, 1, 1, 0,0, 1,1].    
\end{align*} 
The optimal solution is ${\bf{h}} = [1,1]$ with error 3, 
but the solutions [1,0] and [0,1] with error 5 are worse than [0,0] with error 4, and hence [1,1] will not be explored by the greedy algorithm.

\subsubsection{Local Search Strategy}

We now propose a local search strategy to explore around the solution  obtained by the greedy algorithm. 
We limit ourselves to switching at most $q$ entries that are chosen randomly. In our implementation, we choose $2 \leq q \leq \lceil \log r \rceil $, where $\log$ has a base of 2.  
Given a solution $\mathbf{h} \in \{0,1\}^r$, we search randomly in the $q$-radius balls centered on this vector, that is, 
    \[
    B(\mathbf{h}) = \{ \mathbf{s} \in \{0,1\}^r \ | \ \|\mathbf{h}-\mathbf{s}\|_2^2 \leq q \}, 
    \] 
    to check whether there is a better solution. 
    If we obtain a better solution, we recursively call the algorithm on this solution. 

    We set the limit of recursive calls as $T$, and  we used $T=r$ in our implementation.  
    Hence, for $q \leq \lceil \log r \rceil$ and $T=r$, the number of total perturbations is bounded by $r\lceil\log r\rceil$ ($r$ vectors at a distance of $2$, $r$ at a distance of $3$, \dots up to $\lceil\log r\rceil$). Therefore, we have at most $r^2\lceil\log r\rceil$ vectors explored. 
    This gives us a complexity of $\bigO(mr^3\lceil\log r\rceil)$ in the worst case when we consider the comparisons and multiplications between vectors. The algorithm is presented in 
    Algorithm~\ref{LocalSearchAlgo}. We refer to the combination of the greedy algorithm and this local search strategy as \textit{Greedy-BoolLS}. 

\begin{algorithm}[h]
\begin{algorithmic}[1]
\REQUIRE $\mathbf{h} \in \{0,1\}^{r}, \; \mathbf{x}\in \{0,1\}^{m}, \; \mathbf{W} \in \{0,1\}^{m\times r}, \; T \in \mathbb{N}$ \hfill \text{// $T$ is the max number of recursive calls}, $q$ the radius of the search.
\ENSURE $\mathbf{h} \in \{0,1\}^{r}$
\IF{$T = 0$}
    \RETURN $\mathbf{h}$
\ENDIF
\FOR{$i \leftarrow 1,2,\dots,T$ }  
 
    \FOR{$k \leftarrow 2,\dots,q$} 
        \STATE $\mathbf{u} \leftarrow \mathbf{0}_{r}$ \; \hfill \text{// We want to look at another solution vector at distance $k$ from $h$.} 
        \WHILE{$ \sum_j^r u(j) < k$}
            \STATE $idx \leftarrow \mathcal{U}(\{1,2,\dots,r\}) $ \hfill \text{// uniform distribution of the set $\{1,2,\dots,r\}$} 
            \STATE $\mathbf{u}(idx) \leftarrow 1$ \;
        \ENDWHILE
        
        
        \STATE $\textbf{h'} \leftarrow \textbf{h} \oplus \textbf{u}$ (XOR operator) 
        \IF {$\|\textbf{x}-\textbf{W}\circ \textbf{h'}\|_2 < \|\textbf{x}-\textbf{W}\circ \textbf{h}\|_2$}
        \STATE $\textbf{h} \leftarrow \textsc{LocalSearch}(\textbf{h'},\textbf{x},\textbf{W},\;T-1, q)$
        \ENDIF
    \ENDFOR
\ENDFOR
\RETURN $\textbf{h}$ 
\caption{\textsc{LocalSearch}}
\label{LocalSearchAlgo}
\end{algorithmic}
\end{algorithm}
    


\subsection{Heuristic method to combine multiple solutions} 


\textit{Greedy-BLS} can generate different solutions because of the randomized local search.
We now present a greedy heuristic to combine several such solutions. This is a loose adaptation of MS-Comb-AO, where the "building block" algorithm is \textit{Greedy-BLS}. We generate many solutions and store all the rank-one factors in a vectorized form as the columns of a matrix that we call $\mathbf{U}$. 
Let us assume that the number of rank-one factors collected is $N$ and let us also consider a vector $\textbf{y}_2 \in \mathbb{N}^r$. The vector $\textbf{y}_2$ functions similarly to the vector $\textbf{y}$ of MS-Comb-AO (it selects the best rank-one factors among all the collected ones) but is implemented differently. 
The vector $\textbf{y}$ is in $\{0,1\}^N$, where $\textbf{y}_j = 1$ means that the $j$th rank-one factor was chosen, and with the additional constraint that the values of $y$ must sum up to $r$. The vector $\textbf{y}_2$ is in $\mathbb{N}^r$ whose elements designate which rank-one factors we are picking. If, for example, the $j$th rank-one factor is picked, then one of the values of $\textbf{y}_2$ will be equal to $j$. 

We start with the initial best solution (the best rank-$r$ solution among the solutions gathered by the multiple \textit{Greedy-BLS} calls), and try to improve it by randomly switching some of its rank-one factors. The algorithm is very similar to the \textsc{LocalSearch} algorithm. However, rather than applying it vector by vector, it is applied to the rank-one factors. This is shown in Algorithm~\ref{alg:rank1recon} and we refer to it as \textit{Heur-Comb}. 
We refer to the whole algorithm as \textit{Greedy-Comb}, that is, the gathering of solutions through  \textit{Greedy-BLS} and their combination to obtain the best possible $r$ rank-one factors through Algorithm~\ref{alg:rank1recon}. 
Furthermore, similarly to Tree BMF described in  Section~\ref{subsec:hiertreebmf}, we can combine solution in a tree structure, that is, gather solutions from multiple calls to Greedy-Comb, and then combine them with Algorithm~\ref{alg:rank1recon}. 
We refer to this version of the algorithm as \textit{Greedy-TreeBMF}.


    
\begin{algorithm}[h]
        \begin{algorithmic}[1]
            \REQUIRE 
            $\vect(\mathbf{X}) \in \{0,1\}^{mn}$, 
            the columns of $\mathbf{U} \in \{0,1\}^{mn \times N}$ are vectorized rank-one factors, 
            $\mathbf{y}_2 \in [N]^{r}$ collects the $r$ selected rank-one indices,   
            $T_{\max} \in \mathbb{N}$ is the maximum number of improvements, 
            $n$$\_$trials is the number of trials per potential improvement. 
            
            \ENSURE $\mathbf{y}_2 \in [N]^{r}$ 
            \STATE best\_err $= \|\vect(\mathbf{X}) - \sum_{j=1}^r\mathbf{U}(:,\mathbf{y}_2(j))\|_F$ 
            \hfill \text{{// initial error of the selected rank-one factors}}\;
    
            \STATE $T = 1$, iter = 1

            
            \WHILE{iter $\leq $ $n$\_trials  and $T \leq T_{\max}$}
                        \STATE $u\_idx \leftarrow \mathcal{U}(\{1,2,\dots,N\} \backslash \mathbf y_2) $ \; \hfill \text{// pick the index of a column of $U$ not in $\bf y_2$.}
                        \STATE $idx \leftarrow \mathcal{U}(\{1,2,\dots,r\}) $ \; \hfill \text{// pick an index of a column of $U$ in $\bf y_2$}
                        \STATE $\mathbf{y}'_2 = \mathbf{y}_2$, $\mathbf{y}'_2(idx) \leftarrow u\_idx$ 
                        \; \hfill \text{// $\mathbf{y}'_2$ swaps the two above indices.}
                    \IF{$ \| \vect(\mathbf{X}) - \sum_{j=1}^r\mathbf{U}(:,\mathbf{y}'_2(j)) \| <$ best\_err}
                    
                        \STATE $\mathbf{y}_2 \leftarrow \mathbf{y}_2'$ 
                        
                        \STATE best\_err $ = \|\vect(\mathbf{X}) - 
                        \sum_{j=1}^r\mathbf{U}(:,\mathbf{y}_2(j))\|_F $

                        \STATE T = T+1, iter = 1\; 

                    \ELSE 
                    \STATE iter = iter+1\; 
                        
                    \ENDIF
            \ENDWHILE 
            \RETURN $\mathbf{y}_2$
        \caption{Heuristic to combine multiple solutions}    
        \label{alg:rank1recon}
        \end{algorithmic}
    \end{algorithm}

\subsection{New data structure for Boolean vectors and matrices in C++} \label{sec:datastruct} 

The proposed greedy algorithms were written in C++. C++ is a low-level programming language, which means that it allows for control over hardware (for example, directly reading and writing into memory with greater efficiency than in languages like Python or Julia). Although C++ provides a default \texttt{bool} data type, we propose a custom data structure for Boolean matrices that uses less space than a two-dimensional array containing \texttt{bool} elements and can perform operations more quickly. Furthermore, while there are libraries for linear algebra in C++, such as Armadillo \cite{armadillo_C++} or Eigen \cite{eigenweb}, they do not take full advantage of the properties of Boolean algebra.


As a low-level programming language, unlike MATLAB, Python, or Julia, declaring a variable requires a data type as an additional argument. A variable can be, among others, a \texttt{char} (which can store a single character and whose size is 1 byte, i.e., 8 bits), an \texttt{int} (which can store any integer from $-2^{31}$ to $2^{31}$ and whose size is 4 bytes), a \texttt{double} (which can store high precision floating point numbers within the intervals $\pm (2.23*10^{-308},1.797*10^{308})$ and whose size is 8 bytes), or a \texttt{bool} (which takes the values 'true' and 'false' and whose size is 1 byte).
We notice that even in the case of the \texttt{bool} variable, there is a minimum requirement of 8 bits (= 1 byte). Due to Boolean operations only using the values `0' and `1', we create a data structure that solely uses 1 bit to represent entries of Boolean vectors and matrices. 
This allows for considerable improvements in terms of memory usage and computation time for Boolean matrix operations. 

Our design starts by initially creating a custom data structure, which we refer to as a \texttt{bitset} data structure that uses a one-dimensional array. 
If we define an array of \texttt{bool} elements of size $m$, $m$ bytes are used without the ability to manipulate individual bits separately; hence, 8 bits are used to represent a single \texttt{bool} variable. 
With the \texttt{bitset} data structure, we aim to represent each element of the array with only one bit. 
To do so, our \texttt{bitset} data structure uses an \texttt{int} array together with bitwise operations; for example, bitwise-or, bitwise-and, as well as functions that modify or print an element of the array. 
When we define a \texttt{bitset} array with $m$ elements, since each bit corresponds to one element of the array, the size of this custom array is $\lceil \frac{m}{32} \rceil$ elements. If, for example, we define a \texttt{bitset} array of 30 elements, although the user can normally modify and access 30 elements, in reality, it is stored in a single \texttt{int}. 
As an illustration, let us compare a \texttt{bool} array with 50 elements and a \texttt{bitset} array with 50 elements. For the former, we are using 50*(8 bits) = 400 bits. For the latter, the array is of size $\lceil \frac{50}{32} \rceil = 2$. This means that we are using only 64 bits instead of 400 bits. 

To create Boolean matrices in an efficient manner, not just one dimensional arrays, we define another data structure,  \texttt{bitmatrix}, which contains a two dimensional bitset array that we will use to represent Boolean matrices, as well as the definitions of the various operators that we need to use - Boolean matrix multiplication, addition, element access, and element retrieval, to mention a few. 

Since these two data structures are custom, we have to redefine all possible operations from the ground up. Some examples of such operations include accessing an element, changing an element, performing an OR operation or an AND operation between two \texttt{bitset} arrays or \texttt{bitmatrix} data structures. All defined operations are performed by processing individually the bits of the operands. 
For example, the OR operation is implemented as a loop that performs the operation for each bit individually and produces the final result.

\paragraph{Comparison of our \texttt{bitmatrix} data structure} 

Let us present an experiment that shows that our custom implementation outperforms other linear algebra libraries in C++.
 Our setting is the following: We perform Boolean matrix multiplication between two square matrices of size $n \times n$, for $n \in \{100,500,1000,5000,10000,20000\}$, whose entries are rounded from the uniform distribution in $[0,1]$ 
(hence, these matrices have, on average, as many zeros as ones)\footnote{Note that the Boolean product of such matrices will be, with very high probability, the all-one matrix. The goal here is to compare the computational and memory demands of different data structures.}.   
We choose this operation because it is one of the most demanding and ubiquitous matrix operations.
We are comparing with the Eigen and Armadillo C++ linear libraries. From these libraries, the matrices that we are using are of \texttt{float} type, which means that each element represents a real number and uses 32 bits to store it. 

The reason we choose each element to be of \texttt{float} type, instead of \texttt{bool} or \texttt{uint\_8} (where each element would instead be represented by 8 bits), is because of the use of the BLAS routines \cite{BLAS}, which are only available for \texttt{float} and \texttt{double} matrices. BLAS (which stands for "Basic Linear Algebra Subprograms") are routines that provide standard building blocks for performing basic vector and matrix operations. Its matrix multiplication routine, called GEMM (which stands for 'general matrix multiplication'), is heavily optimized and is widely used for dense matrix–matrix multiplication. 

Table~\ref{tab:c++_muls} reports the average time over 10 trials to execute the multiplications (in seconds), for each data size and each data structure considered in Table~\ref{tab:c++_muls}. 
\begin{table}[h]
	\begin{center}
		\begin{tabular}{lllllll}
\hline
			 & 100 & 500 & 1000 & 5000 & 10000 & 20000 \\ [0.5ex] 
			\hline
			 \texttt{bitmatrix}  & $7.21 \times 10^{-5}$  & $\bm{0.0027}$ & $\bm{0.0079}$  & $\bm{0.37}$  & $\bm{3.71}$ & $\bm{40.49}$ \\ 
            Eigen~\cite{eigenweb} & $\bm{ 5.093 \times 10^{-5}} $ & $0.0036$ & $0.0316$  & $3.05$ & $24.88$ & $195.23  $ \\
            Armadillo~\cite{armadillo_C++} & $ 0.00036$  & $0.017$  & $0.047$  & $3.44$  & $26.11$  & $203.86$ \\
            \hline
\end{tabular}
	\end{center}
	\caption{CPU time (in seconds) for Boolean matrix multiplications for six values of the size $n$ and all considered data structures. Reported values are average runtimes over 10 multiplications.}
	\label{tab:c++_muls}
\end{table}

We see that our \texttt{bitmatrix} data structure is faster than the others for all data sizes considered, except for $n = 100$, where Eigen has the best performance. For $n = 500$, \texttt{bitmatrix} is slightly faster than Eigen and much faster than Armadillo. Finally, for all $n \geq 1000$ considered, our \texttt{bitmatrix} data structure is significantly faster than the other two libraries. Especially in the case of $n=20000$, \texttt{bitmatrix} is five times faster than the other two. 

\section{Numerical Experiments}
\label{sec:experiments}
All experiments are performed with a 12th Gen Intel(R) Core(TM) i7-1255U  1.70 GHz, 16GB RAM. In the experiments where we use the methods that solve MIPs, we use Julia v. 1.10. The code is available from \url{https://gitlab.com/ckolomvakis/boolean-matrix-factorization-ip-and-heuristics}. 

\subsection{Small datasets from \cite{BoolMF_IP}}

We now perform experiments on four real binary datasets with no missing data and four real binary datasets with missing data used in~\cite{BoolMF_IP}, which come from~\cite{UCI, krebs2008network}; see Tables~\ref{datasets} and \ref{datasets_incomplete}. As in \cite{BoolMF_IP}, we use $r= 2, 5, 10$ for all datasets. \noindent 
\begin{table}[H]
\begin{center}
\begin{tabular}{lllll}
\hline
    & zoo & heart & lymp & apb \\ 
			\hline
			$m \times n$~~ & $101 \times 17$ & $242 \times 22$ & $148 \times 44$ & $105 \times 105$\\ 
\hline
\end{tabular}
\caption{Four binary real-world datasets.}
	\label{datasets}
\normalsize
\end{center}
\end{table}
\begin{table}[H]
	\begin{center}
\begin{tabular}{lllll}
\hline
    &  tumor & hepatitis & audio
& votes \\
\hline
 $m\times n$ & $339\times24$ & $155\times 38$ & $226\times 92$ &  $435\times 16$ \\
  \#missing & 670 & 334 & 899 & 392\\
            \%observed & 24.3 & 47.2 & 11.3 & 49.2\\
\hline
\end{tabular}
	\end{center}
	\caption{Four binary real-world incomplete datasets.}
	\label{datasets_incomplete}
\end{table}
 In~\cite{BoolMF_IP}, the proposed IP-based approaches for BMF perform well against the state of the art (they used a 20-minute time limit for their method), namely against a greedy scheme~\cite{BoolMF_IP}, 
ASSO and ASSO++~\cite{miettinen2008discrete}, 
a gradient method with a penalty term from~\cite{zhang2007binary}, 
and an NMF-based heuristic. Note that in our conference paper~\cite{KolomvakisMLSP}, we also included results from two recent algorithms based on continuous relaxations, namely~\cite{miron2021boolean, dalleiger2022efficiently}, but they did not provide better results.

Table~\ref{bestres} reports the best results of all these methods for all datasets. From now on, we report the results of our methods by comparing them with the best solution from \cite{BoolMF_IP} provided in Table~\ref{bestres}. That is, for each method, we report the difference between our result and the best result from Table~\ref{bestres}. In particular, this means that a negative value implies that our algorithm found a solution better than all the methods mentioned above.

\begin{table}[ht!]
	\begin{center}
		\begin{tabular}{lccc}
\hline
			& $r = 2$ ~ & $r = 5$ ~ & $r = 10$ ~\\ [0.5ex] 
            \hline
			\hline
			zoo &  271 & 	126	 & 39\\
			\hline 
			heart &  1185 &	737 &	419 \\
			\hline
			lymp &  1184	& 982 &	728 \\ 
			\hline
			apb &  776 &	684	& 573\\
			\hline
			\hline
            tumor &  1352 & 962	 & 514\\
			\hline 
			hepatitis & 1264 &	1138 & 907 \\
			\hline
			audio &  1419 & 1064 & 765 \\ 
			\hline
			votes &  1246 &	779	& 240\\
			\hline
\end{tabular}
	\end{center}
	\caption{Objective function 
 $\|\mathbf{M} \odot (\mathbf{X}-\min(1,\mathbf{WH}))\|_F^2$ of the best solution found by various algorithms in~\cite[Table~4, page~20]{BoolMF_IP}. 	\label{bestres}} 
\end{table}

For each of our algorithms, we consider two time limits: $T = 30$ seconds and $T = 5$ minutes, hence smaller than the 20-minute time limit used in~\cite{BoolMF_IP}. 
Our algorithms are the following:  
\begin{itemize}
    \item \textbf{MS-AO:} We generate as many BMFs as possible with AO-BMF within the time $T$ and return the best solution. The initialization of AO-BMF is chosen alternatively as one of the two strategies (NMF-based or random columns/rows of $\bf{X}$).  
    \item \textbf{MS-Comb-AO:} As explained in Section~\ref{sec:Comb}, we generate as many BMFs as possible with AO-BMF within time $3T/4$ and then combine them by solving~\eqref{Comb_BoolMF} with a time limit of $T/4$. We used the same random seed as MS-AO so that the solutions generated are the same, except that fewer solutions are generated, since only 3/4 of the total time is spent. 
    \item \textbf{Tree BMF:} We consider the depth of the tree to be 1. For $T = 30$ seconds, the leaf nodes will gather 5 solutions, while for $T = 5$ minutes, they will gather 15 solutions. When moving to the next level of the tree, the given time to compute a BMF is divided by two. Furthermore, in a non-leaf node, the combining step is also performed in $T/2$ time.
     \item \textbf{Greedy-Comb:} For both $T = 30$ seconds and $T = 5$ minutes, we gather solutions using Greedy-BoolLS for $T$ seconds to fill the $\mathbf{U}$ matrix. We repeat 5 trials and report the best solution.
    \item \textbf{Greedy-TreeBMF:} We gather the solutions from several Greedy-Comb calls and greedily pick the final solution among them. For $T = 30$ seconds, Greedy-Comb is called 3 times, and each call has a limit of $10$ seconds. For $T = 5$ minutes, Greedy-Comb is called 5 times, and each call has a limit of $60$ seconds to fill the $\mathbf{U}$ matrix.  We report the best results among 5 trials.
\end{itemize}

\paragraph{Results and discussion} 

Tables \ref{complete_all_times} and \ref{incomplete_all_times} 
 report the results.  
\begin{table}[ht!]
	\begin{center}
		\begin{tabular}{ c |c |c | c | c | c |c} 
			&  & MS-AO & MS-Comb-AO & \makecell{Tree \\BMF} & \makecell{Greedy \\Comb} & \makecell{Greedy \\TreeBMF} \\ [0.5ex] 
			\hline
\multirow{4}{3em}{$r=2$} & zoo & \hspace{1.1mm} \textbf{0} $ \ | \ $ \textbf{0} & \textbf{0} $ \ | \ $ \textbf{0} & \textbf{0} $ \ | \ $ \textbf{0} & \textbf{0} $ \ | \ $ \textbf{0} & \textbf{0} $ \ | \ $ \textbf{0}\\ 
& heart & \; \underline{+2} $ \ | \ $  \underline{+2}  & \hspace{-4mm} \underline{+2} $ \ | \ $ \textbf{0} & \underline{+2} $ \ | \ $ \underline{+2} & \underline{+2} $ \ | \ $ \underline{+2} & \underline{+2} $ \ | \ $ \underline{+2}\\
& lymp &  \ \  \textbf{-10}  $ \ | \ $ \textbf{-10} & \textbf{-10} $ \ | \ $ \textbf{-10} & \textbf{-10} $ \ | \ $ \textbf{-10} & \textbf{-10} $ \ | \ $ \textbf{-10} & \underline{-7} $ \ | \ $ \underline{-7}\\
& apb & \ \textbf{0} \ $ \ | \ $ \textbf{0} & \textbf{0} $ \ | \ $ \textbf{0} &  \textbf{0} $ \ | \ $ \textbf{0} & \textbf{0} $ \ | \ $ \textbf{0} & \textbf{0} $ \ | \ $ \textbf{0} \\ 
			\hline
\multirow{4}{3em}{$r=5$} & zoo & \  \ \textbf{-1} $ \ | \ $  \textbf{-1}  & \textbf{-1} $ \ | \ $ \textbf{-1} &  \textbf{-1} $ \ | \ $ \textbf{-1} &  \textbf{-1} $ \ | \ $ \textbf{-1} & \textbf{-1} $ \ | \ $ \textbf{-1}\\ 
&heart & \ \textbf{-1} \ $ \ | \ $ \textbf{-1} & \textbf{-1} $ \ | \ $ \textbf{-1}  & \textbf{-1} $ \ | \ $ \textbf{-1} & \textbf{-1} $ \ | \ $ \textbf{-1} & \textbf{-1} $ \ | \ $ \textbf{-1} \\
&lymp &  \ \  -25  $ \ | \ $ -32  & -25 $ \ | \ $ -32 & -27 $ \ | \ $ \underline{-34} & \textbf{-38} $ \ | \ $ \textbf{-38} & \hspace{-0.45mm} -33 $ \ | \ $ \textbf{-38}\\
&apb & \hspace{-0.4mm} +6 $ \ | \ $ \underline{-6}  & -1 $ \ | \ $ \underline{-6}  & \hspace{-0.65mm} \underline{-6} $ \ | \ $ \textbf{-7} & \underline{-6} $ \ | \ $ \underline{-6} & \hspace{-1.55mm} \textbf{-7} $ \ | \ $ \underline{-6}\\
\hline
\multirow{4}{3em}{$r=10$} & zoo & \hspace{-1.45mm} \underline{+3}  $ \ | \ $ \textbf{0}  & \textbf{0} $ \ | \ $ \textbf{0} & \textbf{0} $ \ | \ $ \textbf{0} & +7 $ \ | \ $ \underline{+3} & +8 $ \ | \ $ +6\\ 
&heart & \hspace{1.15mm} \textbf{0} $ \ | \ $ \textbf{0} & \textbf{0} $ \ | \ $ \textbf{0} & \textbf{0} $ \ | \ $ \textbf{0} & \underline{+18} $ \ | \ $ \underline{+18} & \underline{+18} $ \ | \ $ \underline{+18}\\
&lymp &  \hspace{1.9mm} -15  $ \ | \ $  \textbf{-34}  & \hspace{-0.55mm} -15 $ \ | \ $ \textbf{-34} & \hspace{-0.55mm} -19 $ \ | \ $ \textbf{-34} & -12 $ \ | \ $ -24 &\ \hspace{-0.6mm} -9 $ \ | \ $ \underline{-31}\\
&apb & \ +4 \ $ \ | \ $ +2 & +2 \: $  | \ $ \textbf{-7} \: & \textbf{-7} $ \ | \ $ \textbf{-7} & -1 $ \ | \ $ -2 & -2 $ \ | \ $ \underline{-4}\\ 
        	\hline
		\end{tabular}
	\end{center}
	\caption{Results for all our proposed methods on the datasets from Table~\ref{datasets} for two  time limits: 30 seconds (left) and 5 minutes (right). 
    Bold means the best solution found among our algorithms, underline is the second best.}   
	\label{complete_all_times}
\end{table}   
\begin{table}[ht!]
	\begin{center}
		\begin{tabular}{ c |c |c | c |c|c|c} 
			&  & MS-AO & MS-Comb-AO & \makecell{Tree \\BMF} & \makecell{Greedy \\Comb} & \makecell{Greedy \\TreeBMF}\\ [0.5ex] 
			\hline
\multirow{4}{3em}{$r=2$} & tumor & \hspace{-0.02mm} \begin{tabular}{@{}c|c@{}}
    \hspace{-0.1mm} \underline{+2}  & \textbf{-1} \;
    \end{tabular}  & \begin{tabular}{@{}c|c@{}}
      \textbf{-1} & \textbf{-1}
    \end{tabular} & \begin{tabular}{@{}c|c@{}}
      \textbf{-1} & \textbf{-1}
    \end{tabular} & \begin{tabular}{@{}c|c@{}}
      \textbf{-1} & \textbf{-1}
    \end{tabular} & \begin{tabular}{@{}c|c@{}}
      \textbf{-1} & \textbf{-1}
    \end{tabular}\\ 
& hep/tis & \ \begin{tabular}{@{}c|c@{}}
      \textbf{0} & \textbf{0}
    \end{tabular}  & \begin{tabular}{@{}c|c@{}}
      \textbf{0} & \textbf{0}
    \end{tabular} & \begin{tabular}{@{}c|c@{}}
      \textbf{0} & \textbf{0}
    \end{tabular} & \begin{tabular}{@{}c|c@{}}
      \textbf{0} & \textbf{0}
    \end{tabular} & \begin{tabular}{@{}c|c@{}}
      \textbf{0} & \textbf{0}
    \end{tabular}\\
& audio & \ \begin{tabular}{@{}c|c@{}}
      \textbf{-8} & \textbf{-8}
    \end{tabular} & \begin{tabular}{@{}c|c@{}}
      \textbf{-8} & \textbf{-8}
    \end{tabular} & \begin{tabular}{@{}c|c@{}}
      \textbf{-8} & \textbf{-8}
    \end{tabular} & \begin{tabular}{@{}c|c@{}}
      \textbf{-8} & \textbf{-8}
    \end{tabular} & \begin{tabular}{@{}c|c@{}}
      \textbf{-8} & \textbf{-8}
    \end{tabular}\\
& votes & \ \begin{tabular}{@{}c|c@{}}
      \textbf{0} & \textbf{0}
    \end{tabular} & \begin{tabular}{@{}c|c@{}}
      \textbf{0} & \textbf{0}
    \end{tabular} & \begin{tabular}{@{}c|c@{}}
      \textbf{0} & \textbf{0}
    \end{tabular} & \begin{tabular}{@{}c|c@{}}
      \textbf{0} & \textbf{0}
    \end{tabular} & \begin{tabular}{@{}c|c@{}}
      \textbf{0} & \textbf{0}
    \end{tabular}\\ 
			\hline
\multirow{4}{3em}{$r=5$} & tumor & \begin{tabular}{@{}c|c@{}}
      +17 & \; -7
    \end{tabular}  & \begin{tabular}{@{}c|c@{}}
    \:  -3 \hspace{-0.3mm} & -10 \hspace{-0.05mm}
    \end{tabular} &  \begin{tabular}{@{}c|c@{}}
      \underline{-11} & -10
    \end{tabular} &  \begin{tabular}{@{}c|c@{}}
      \hspace{0.45mm} -5 & \textbf{-19} \hspace{-2.23mm}
    \end{tabular} &  \begin{tabular}{@{}c|c@{}}
      \hspace{1.25mm} -1 & \textbf{-19}
    \end{tabular}\\ 
& hep/tis &  \begin{tabular}{@{}c|c@{}}
     \ -119 & -139
    \end{tabular} & \begin{tabular}{@{}c|c@{}}
      -120 & -129
    \end{tabular} & \begin{tabular}{@{}c|c@{}}
      -128 & -135
    \end{tabular} & \begin{tabular}{@{}c|c@{}}
      -138 & \underline{-143}
    \end{tabular} & \begin{tabular}{@{}c|c@{}}
     -136 & \textbf{-148}  \hspace{-2.2mm}
    \end{tabular}\\
&audio &  \begin{tabular}{@{}c|c@{}}
     \ -24 & -25
    \end{tabular}  & \begin{tabular}{@{}c|c@{}}
      -17 & -25
    \end{tabular} & \begin{tabular}{@{}c|c@{}}
      -25 & -25
    \end{tabular} & \begin{tabular}{@{}c|c@{}}
      \hspace{-0.5mm} -26 & \textbf{-29}
    \end{tabular} & \begin{tabular}{@{}c|c@{}}
      \hspace{0.55mm} -9 & \underline{-27}
    \end{tabular}\\
&votes &  \begin{tabular}{@{}c|c@{}}
      \hspace{-2.2mm} \: +35 & -12 \hspace{0.3mm} 
    \end{tabular}  & \begin{tabular}{@{}c|c@{}}
       +27 &\ -8 \;
    \end{tabular}  & \begin{tabular}{@{}c|c@{}}
      \hspace{-0.25mm} -60 & \textbf{-78} \hspace{-1.0mm}
    \end{tabular} & \begin{tabular}{@{}c|c@{}}
      \hspace{-0.75mm} \underline{-66} & \underline{-66} \hspace{-0.75mm}
    \end{tabular} & \begin{tabular}{@{}c|c@{}}
      \hspace{-1.15mm} \underline{-66} & \underline{-66} \hspace{-1.05mm}
    \end{tabular}\\
\hline
\multirow{4}{3em}{$r=10$} & tumor & \ \begin{tabular}{@{}c|c@{}}
      +68 & +18
    \end{tabular}  & \begin{tabular}{@{}c|c@{}}
      \underline{-2}  \hspace{-0.01mm} & \textbf{-4} \hspace{-0.35mm}
    \end{tabular} & \begin{tabular}{@{}c|c@{}}
      \hspace{-0.4mm} \textbf{-4} & \underline{-2} \hspace{-0.15mm}
    \end{tabular} & \begin{tabular}{@{}c|c@{}}
      +15 \hspace{-1.75mm} & +8 \hspace{-0.15mm}
    \end{tabular} & \begin{tabular}{@{}c|c@{}}
      \hspace{-0.5mm} +3 & +3 \hspace{-0.15mm}
    \end{tabular}\\ 
& hep/tis &  \begin{tabular}{@{}c|c@{}}
    \  -113 & -145
    \end{tabular} & \begin{tabular}{@{}c|c@{}}
      \hspace{-0.05mm} -107 & -151 \hspace{0.15mm}
    \end{tabular}  & \begin{tabular}{@{}c|c@{}}
      \hspace{-0.01mm} -145 & -154 \hspace{-0.25mm}
    \end{tabular}  & \begin{tabular}{@{}c|c@{}}
      \hspace{0.05mm} -147  & \underline{-157} \hspace{-0.25mm}
    \end{tabular}  & \begin{tabular}{@{}c|c@{}}
      \hspace{0.4mm} -146 & \textbf{-158} \hspace{-0.35mm}
    \end{tabular}\\
&audio &  \ \  \begin{tabular}{@{}c|c@{}}
      +16 & +16 \
    \end{tabular}  & \begin{tabular}{@{}c|c@{}}
      +2 & \underline{-12}
    \end{tabular} & \begin{tabular}{@{}c|c@{}}
    \;  -3 \hspace{-0.2mm} & \textbf{-16} \hspace{-0.75mm}
    \end{tabular} & \begin{tabular}{@{}c|c@{}}
    \; \hspace{-0.2mm} +41 & +39 \hspace{1.85mm}
    \end{tabular} & \begin{tabular}{@{}c|c@{}}
    \; \hspace{-0.2mm} +37 & +30 \hspace{2.45mm}
    \end{tabular}\\
&votes & \ \begin{tabular}{@{}c|c@{}}
      +295 & +213
    \end{tabular} & \begin{tabular}{@{}c|c@{}}
      +135 & +104 \hspace{-1.08mm}
    \end{tabular} & \begin{tabular}{@{}c|c@{}}
      \hspace{-0.01mm} +153 & +52 \hspace{1.65mm}
    \end{tabular} & \begin{tabular}{@{}c|c@{}}
      \hspace{-0.01mm} +20 & \underline{-4} \hspace{3.35mm}
    \end{tabular} & \begin{tabular}{@{}c|c@{}}
      \hspace{0.41mm} +27 & \textbf{-15} \hspace{1.65mm}
    \end{tabular}\\ 
        	\hline
		\end{tabular}
	\end{center}
	\caption{Results for all our proposed methods on the datasets from Table~\ref{datasets_incomplete} with missing entries for two time limits: 30 seconds (left) and 5 minutes (right). 
    Bold means the best solution found among our algorithms, underline is the second best.}   
	\label{incomplete_all_times}
\end{table}
 
 \noindent Before going into the details, we observe that most entries of the tables are negative, meaning that our proposed algorithms outperform the state of the art in most cases. 
 Let us discuss each of our algorithms separately: 
\begin{itemize}
    \item \textbf{MS-AO:} Quite surprisingly, MS-AO is already able to perform on par with or improve upon the state of the art. With a 30 seconds time limit, it does on 14 out of 24 cases: 9/24 cases show improvements, sometimes significant, as for the hepatitis dataset (-119 for $r=5$ and -113 for $r=10$). 
    With a 5 minutes time limit, it performs better or on par on 19 out of the 24 cases. It is only significantly worse for $r=10$ for incomplete datasets. 

    \item \textbf{MS-Comb-AO:} For some datasets, AO is already able to generate very good solutions, and hence solving~\eqref{Comb_BoolMF} is not useful, e.g., for the lymp dataset.  
    However, in most cases, this combination is beneficial, sometimes significantly. In particular, with the time limit of 30 seconds on the apb dataset for $r=5$, the best solution found by MS-AO has an error of 689 
    (+6), while the combination leads to an error of 682 
    (-1); for the zoo dataset with $r=10$, 
    it goes from 42 (+3) to 39 (0). A similar behaviour is observed for 5 minutes: for the apb dataset with $r=10$, it changes from +2 to -7, and forthe heart dataset, it changes from +2 to 0. 

    MS-Comb-AO, with a 5-minute time limit, is able to perform on par with or better than the state of the art in 23 out of 24 cases, sometimes providing significantly better solutions (e.g., for the hepatitis dataset with -129 for $r=5$ and -151 for $r=10$).
    
   
    \item \textbf{Tree-BMF:} 
    It provides improvement over MS-Comb-AO, sometimes significantly, except for the "heart" dataset with $r = 2$. 
    In fact, Tree-BMF provides the best solution in 15 out of 24 cases. 
    
    \item \textbf{Greedy-Comb:} 
    In 19 out of 24 cases, Greedy-Comb improves or performs on par with the state of the art. 
    Moreover, for the datasets with missing entries, it provides the best solution in 12/24 cases. 
    
    \item \textbf{Greedy-TreeBMF: } It improves upon Greedy-Comb. In particular, it provides the best solution in 13/24 cases.  
\end{itemize}

\paragraph{Summary}  For the majority of datasets, our methods either perform on par with or better than the state of the art (Table~\ref{bestres}), sometimes by a significant margin. 
For all datasets, at least one of our methods performs on par with or  improves upon the state of the art. 

The greedy-based methods generally yield competitive results (with some exceptions, namely for the `heart' dataset for $r = 2$ and $r = 10$, and for the `hepatitis' dataset for $r = 10$.), despite being less complex than the IP-based methods. 
There are even some cases where the greedy-based methods outperform the IP-based ones, e.g., for the `votes' and `hepatitis' datasets. 


\subsection{Topic modeling}  
\label{subsec:topic_modelling} 

In this section, we consider the NIST tdt2 dataset, which contains news stories from 1998 across 30 different topics \cite{tdt_reference}. 
The input data matrix, $\mathbf{X}$, consists of 9394 documents with 19528 words. 
In order to test the IP-based methods and to discard unimportant words and documents, we work with subsets of the original dataset that we create as follows: We perform NMF on the original dataset, $\bf X \approx WH$, with rank $r_{\text{NMF}}$, obtain the $w$ most frequent words per topic from the $\mathbf{W}$ factor, and the $d$ most frequent documents per topic from the $\mathbf{H}$ factor. We then consider the corresponding submatrix of $\mathbf{X}$ containing only these words and documents, denoted by $\hat{\mathbf{X}}$.
We later binarize the matrix by assigning all nonzero values to 1. We denote this binarized matrix as $\hat{\mathbf{X}}_b$. We consider three subsets; see Table \ref{tab:Xb_matrices}. 
\begin{table}[h]
	\begin{center}
		\begin{tabular}{lllll}
			\hline
			& $r_{\text{NMF}}$ ~ & $w$ ~ & $d$ ~ & \makecell{Dimensions of $\hat{\mathbf{X}}_b$} \\ [0.5ex] 
			\hline
			small dataset & 20 & 20 & 50 & $302 \times 917$\\
			\hline 
			medium dataset & 20 & 150 & 30 & $1608 \times 517$\\
			\hline
			large dataset & 30 & 1300 &	800 & $10729 \times 8200$ \\ 
			\hline
		\end{tabular}
	\end{center}
	\caption{The three subsets of the NIST tdt2 dataset.}  
	\label{tab:Xb_matrices}
\end{table}\vspace{-1.2mm}

\paragraph{Compared algorithms for the larger datasets}

In this section and the next, we compare our algorithms to the following: 
\begin{itemize}
    \item \textbf{ELBMF} \cite{dalleiger2022efficiently}. An algorithm that uses proximal gradient with both an $l_1$ and an $l_2$ regularizing term. We chose this algorithm because it is a recent one that falls into the category of continuous algorithms for BMF. As we will see later, it scales very well.
    
    \item \textbf{Methods in \cite{Avellaneda_Villemaire_2022}. } This paper  searches for undercover/"from-below" approximations for BMF and uses a MAXSat encoding to solve the optimal Boolean k-undercover problem. MaxSAT is the optimization version of the Boolean Satisfiability Testing problem.
    We compare two methods from the paper: \texttt{FastUndercover}, which is a greedy algorithm, and \texttt{optiblock*} , which uses \texttt{FastUndercover} as an initialization. Even though these methods emphasise avoiding the types of errors that we previously mentioned, the experiments in \cite{Avellaneda_Villemaire_2022} show that they are comparable to other state-of-the-art algorithms that solve BMF.
    \item \textbf{ASSO \cite{miettinen2008discrete}.} As a reference, we are also comparing with the seminal ASSO algorithm \cite{miettinen2008discrete}.
\end{itemize}
We selected the methods from \cite{dalleiger2022efficiently} and \cite{Avellaneda_Villemaire_2022} because they are recent and rely on substantially different modeling approaches. We also included ASSO as a standard baseline. In contrast, we did not consider the IP-based method from \cite{BoolMF_IP}, as it does not scale adequately to the problem sizes studied here. For the methods evaluated, we use the following parameter settings:
\begin{itemize}
    \item \textbf{IP-based methods.}     
    For MS-AO, we report the best results among 10 trials while, for $r=10$, we set a (soft) time limit of $T = 60$ seconds and the number of solutions to 10, while for $r= 20$ we set a limit of $T = 600$ seconds and 10 solutions. 
    For MS-Comb-AO and Tree-BMF, we report the best result among 10 trials as well. For Tree-BMF, for $r = 10$ we consider 3 child nodes, a time limit of $T = 1000$ seconds, 10 solutions for the leaf nodes and depth equal to 1. For $r = 20$, we consider two child nodes, a time limit of $T = 600$ seconds and a depth of 1 for the tree. 
    
    \item \textbf{Greedy-Comb and Greedy-TreeBMF.} 
    For Greedy-Comb, for all instances we gather 10 solutions, with the exception of the large dataset and for $r = 20$, for which we gather 15 solutions. We report the best result over 10 experiments. For Greedy-TreeBMF, we consider the following parameters:
    \begin{itemize}
        \item For the small dataset, for $r = 10$, we set a time limit of 8 seconds and we make 2 calls to Greedy-Comb, each with a 4 second limit. For $r = 20$, we set a time limit of 24 seconds and we make 4 calls to Greedy-Comb, each having a time limit of 6 seconds.
        \item For the medium dataset, for $r = 10$, we set a total time limit of 30 seconds and call Greedy-Comb 5 times with a six second time limit for each Greedy-Comb call. For $r = 20$, we set a total time limit of 80 seconds and call Greedy-Comb 4 times with a 20 second time limit for each Greedy-Comb call. 
        \item For the large dataset, for $r = 10$ the timelimit is set to 1000
        seconds, we make 5 Greedy-Comb function calls and each gathering Greedy-Comb algorithm is given a timelimit of 200 seconds. For $r = 10$ the timelimit is set to 3000 seconds, we make 5 Greedy-Comb function calls and each gathering Greedy-Comb algorithm is be given a timelimit of 600 seconds.
    \end{itemize}
    \item \textbf{ELBMF \cite{dalleiger2022efficiently}.} 
    The parameters used are the ones provided by the authors: \texttt{l1\_reg} = \texttt{l2\_reg} = 0.01, \texttt{maxiter} = 100, \texttt{tol} = $10^{-10}$, \texttt{beta} = $10^{-4}$ and regularization rate $\lambda = 1.0133$. 
    For the remainder of this paper, the number of Monte Carlo trials is 15 and we report the best result among them. 
    
    \item \textbf{Methods in \cite{Avellaneda_Villemaire_2022}.} We report the best result over 10 trials. 
    \item \textbf{ASSO \cite{miettinen2008discrete}.} We report the best result over 10 trials. 
    Furthermore, ASSO requires a threshold parameter that is in $[0,1]$ to search for association among columns in the input matrix. Through trial and error, we find a good threshold parameter for each of the datasets considered: 
    \begin{itemize}
        \item For the small dataset, the threshold is equal to 0.4.
        \item For the medium and large datasets, the threshold is equal to 0.3.
    \end{itemize}
\end{itemize}

\noindent Furthermore, for all the tests, we set an upper time limit of 2 days, after which we terminate the method.  

\paragraph{Results} 

Table~\ref{tab:documents_all} reports the best relative error $\frac{\| \hat{\mathbf{X}}_b - \bf W \circ H \|_F}{\| \hat{\mathbf{X}}_b\|_F}$, in percent, obtained by the different algorithms. 
\begin{table}[h]
  \centering
  \begin{tabular}{l|cc|cc|cc}
    \hline
    & \multicolumn{2}{c|}{Small} 
    & \multicolumn{2}{c|}{Medium} 
    & \multicolumn{2}{c}{Large} \\
    \cmidrule(lr){2-3} \cmidrule(lr){4-5} \cmidrule(lr){6-7}
    Algorithm
    & $r=10$ & $r=20$ 
    & $r=10$ & $r=20$ 
    & $r=10$ & $r=20$ \\
    \midrule
    MS-AO                    & \makecell{84.54\% \\ (83s)} & \makecell{78.19\% \\ (126s)} & \makecell{\underline{94.05\%} \\ (224s)} & \makecell{\textbf{90.71\%} \\ (364s)} & - & - \\
    \hline
    MS-Comb-AO                & \makecell{\underline{84.36\%} \\ (358s)} & \makecell{\underline{78.13\%} \\ (583s)} & - & - & - & - \\
    \hline
    Tree-BMF              & \makecell{\textbf{84.31\%} \\ (781s)} & \makecell{\textbf{78.12\%} \\ (611s)} & - & - & - & - \\
    \hline
    Greedy-Comb               & \makecell{84.94\% \\ (7s)} & \makecell{79.27\% \\ (24s)} & \makecell{94.24\% \\ (17s)} & \makecell{91.01\% \\ (80s)} & \makecell{\textbf{98.56\%} \\ (989s)} & \makecell{\underline{98.11\%} \\ (7497s)} \\
    \hline
    Greedy-TreeBMF            & \makecell{84.61\% \\ (11s)} & \makecell{79.25\% \\ (37s)} & \makecell{\textbf{93.93\%} \\ (41s)} & \makecell{\underline{90.96\%} \\ (112s)} & \makecell{98.68\% \\ (1786s)} & \makecell{98.45\% \\ (5163s)} \\
    \hline
    FastUndercover \cite{Avellaneda_Villemaire_2022}           & \makecell{98.43\% \\ (2.6s)} & \makecell{97.66\% \\ (4.5s)} & \makecell{98.8\% \\ (18s)} & \makecell{98.81\% \\ (12s)} & - & - \\
    \hline
    OptiBlock* \cite{Avellaneda_Villemaire_2022}               & \makecell{97.87\% \\ (2249s)} & \makecell{96.5\% \\ (5869s)} & \makecell{98.52\% \\ (18573s)} & - & - & - \\
    \hline
    ELBMF \cite{dalleiger2022efficiently} & \makecell{91.5\% \\ (0.28s)} & \makecell{88.93\% \\ (0.36s)} & \makecell{95.08\% \\ (0.68s)} & \makecell{94.1\% \\ (0.95s)} & \makecell{99.23\% \\ (67s)} & \makecell{98.29\% \\ (74s)} \\
    \hline
    ASSO \cite{miettinen2008discrete} & \makecell{85.43\% \\ (0.64s)} & \makecell{82.26\% \\ (1.2s)} & \makecell{94.64\% \\ (1.05s)} & \makecell{93.04\% \\ (1.9s)} & \makecell{\underline{98.66\%} \\ (333s)} & \makecell{\textbf{98.10\%} \\ (686s)} \\
    \bottomrule
  \end{tabular} 
  \caption{Relative errors for the datasets from Table~\ref{tab:Xb_matrices}. In parenthesis is the mean time among all Monte Carlo trials (if multiple trials are performed for a method). Bold indicates the best performing method, underline is the second best. 
  The symbol `-' means that the method could not run, because the memory requirements were too demanding.   \label{tab:documents_all}}    
\end{table}

What we immediately notice is that the methods in \cite{Avellaneda_Villemaire_2022} are noticeably behind the other methods. Among all the methods we compare, ASSO appears, on average, to perform very well and to be fast. Our IP-based methods generally perform the best in terms of reconstruction error. 
However, these methods could not be tested on all datasets; in particular, MS-Comb-AO and Tree-BMF could only run on the small dataset. Our Greedy methods are performing well, being on average a close second, in terms of error, to our IP-based algorithms.  ELBMF performs better than FastUndercover and Optiblock*, but it generally trails our proposed methods and ASSO. 

For the instances where we could not report results for our Gurobi-based methods, the reason is that the runs exceeded the available memory.
For the methods in \cite{Avellaneda_Villemaire_2022}, the instances for which we do not report results are those where the computations exceeded two days of runtime.

\subsubsection{Extracting topics with BMF}  

Let $\mathbf{W}$ and $\mathbf{H}$ be a BMF of $\bf X$. 
Since $\bf X$ is a word-by-document matrix, 
\begin{itemize}
    \item the columns of $\mathbf{W}$ correspond to topics: $\mathbf{W}(i,k) = 1$ if the word $i$ belongs to topic $k$, and 

    \item the rows of $\bf H$ indicate which documents belong to which topic: $\mathbf{H}(k,j) = 1$ if document $j$ discusses topic $k$.  

\end{itemize}
Equivalently, each topic is associated with a set of words and a set of documents, indicated by the rank-one binary matrix $\mathbf{W}(:,k)\mathbf{H}(k,:)$. 
However, BMF does not provide word importances within a topic. 
 To retrieve the importance of a word in a topic, we count how many times it is used by the documents of that topic; this is given by the formula 
\begin{equation} \label{eq:Wt}
    \mathbf{W}_t = \mathbf{W} \odot ({\mathbf{X}}\mathbf{H}^\top), 
\end{equation} 
where $\mathbf{W}_t(i,k)$ is equal to zero if the $i$th word does not belong to the $k$th topic; otherwise, it is the number of times the $i$th word is used by the documents belonging to the $k$th topic. 

To show that BMF extracts meaningful topics from a large dataset, Tables~\ref{tab:greedy_comb_words1}-\ref{tab:greedy_comb_words2}-\ref{tab:greedy_comb_words3}  
provide the 10 most frequent words for the 20 topics extracted by Greedy-Comb. 
We observe that these topics are meaningful and can be easily interpreted. 
This solution was obtained by using a specific procedure that allows Greedy-Comb to extract more diverse topics; see Appendix~\ref{app:enhanceprocedure}.

\begin{table}[ht!]
    \centering
    \begin{tabular}{lllllll} 
         \hline
         \makecell{Clinton \\ Lewinsky \\ scandal} & \makecell{Israel \\ Palestine\\ conflict} & \makecell{Ted\\ Kaczynski \\ trial} & \makecell{Indian\\ Pakistani \\ nuclear\\ tests} & \makecell{1998\\ Superbowl} & \makecell{Southeast\\ Asian  econ. \\crisis} & \makecell{1998 US \\ Iraq crisis}\\
         \hline \hline
        `president' & `israel' & `kaczynski' & `nuclear' & `denver' & `economic' & `iraq'\\
        `lewinsky' & `netanyahu' & `lawyers' & `india' & `packers' & `percent' & `weapons'\\
        `clinton' & `israeli' & `defense' & `pakistan' & `super' & `crisis' & `united'\\
        `house' & `palestinian' & `trial' & `tests' & `game' & `government' & `iraqi'\\
        `white' & `peace' & `judge' & `indias' & `bowl' & `indonesia' & `inspectors'\\
        `starr' & `arafat' & `kaczynskis' & `test' & `broncos' & `asian' & `annan'\\
        `lawyers' & `palestinians' & `case' & `weapons' & `green' & `economy' & `council'\\
        `jones' & `talks' & `prosecutors' & `indian' & `bay' & `billion' & `baghdad'\\
        `case' & `minister' & `burrell' & `united' & `elway' & `financial' & `saddam'\\
         \hline
    \end{tabular}
    \caption{Topics 1 - 7 extracted by Greedy-Comb on the large document dataset.}
    \label{tab:greedy_comb_words1}
\end{table}
\vspace{-0.5mm}
\begin{table}[ht!]
    \centering
    \begin{tabular}{lllllll} 
         \hline
         \makecell{Jonesboro \\ school\\ shooting} & \makecell{Tobacco\\ master \\ settlement\\ agreement} & \makecell{Winter \\ Olympic\\ games} & \makecell{Easing of  \\ US embargo\\ of Cuba} & \makecell{Court martial\\ of  Gene\\ McKinney} & \makecell{1998 \\Cavalese\\ cable car\\ crash} & \makecell{1998 NBA\\ finals}\\
         \hline \hline
         `school' & `tobacco' & `olympic' & `cuba' & `mckinney' & `italian' & `game'\\
         `students' & `industry' & `nagano' & `cuban' & `sergeant' & `italy' & `bulls'\\
        `boys' & `bill' & `olympics' & `castro' & `sexual' & `marine' & `jordan'\\
        `jonesboro' & `companies' & `games' & `embargo' & `major' & `crew' & `jazz'\\
        `middle' & `legislation' & `gold' & `american' & `army' & `plane' & `malone'\\
        `teacher' & `smoking' & `medal' & `visit' & `court' & `cable' & `chicago'\\
        `mitchell' & `congress' & `team' & `government' & `gene' & `military' & `points'\\
        `shooting' & `senate' & `won' & `clinton' & `women' & `accident' & `pippen'\\
        `fire' & `tax' & `japan' & `flights' & `martial' & `flying' & `left'\\
    \hline
    \end{tabular}
    \caption{Topics 8 - 14 extracted by Greedy-Comb on the large document dataset.}
    \label{tab:greedy_comb_words2}
\end{table}
\vspace{-0.5mm}
\begin{table}[ht!]
    \centering
    \begin{tabular}{llllll} 
         \hline
         \makecell{Death of \\MLK's assassin } & \makecell{ Visit of the pope \\to Cuba} & \makecell{Execution of \\a US convict } & \makecell{Lawsuit against \\ Oprah Winfrey } & \makecell{Indian\\elections } & \makecell{ Algerian \\ civil war}\\
         \hline \hline
        `ray' & `cuba' & `death' & `winfrey' & `party' & `algeria'\\
        `king' & `pope' & `tucker' & `texas' & `india' & `government'\\
        `james' & `cuban' & `penalty' & `oprah' & `government' & `algerian'\\
        `earl' & `castro' & `texas' & `show' & `hindu' & `islamic'\\
        `martin' & `visit' & `execution' & `cattle' & `congress' & `algiers'\\
        `luther' & `church' & `executed' & `beef' & `bjp' & `killed'\\
        `assassination' & `john' & `row' & `disease' & `election' & `violence'\\
        `kings' & `paul' & `clemency' & `cow' & `seats' & `militants'\\
        `family' & `havana' & `woman' & `amarillo' & `indias' & `armed'\\
         \hline
    \end{tabular}
    \caption{Topics 15 - 20 extracted by Greedy-Comb on the large document dataset.}
    \label{tab:greedy_comb_words3}
\end{table}

\subsection{Facial images}\label{subsec:faces_exp}

We now consider the CBCL facial image dataset. Each column of the data matrix, $\mathbf{X} \in \mathbb{R}^{361 \times 2429}$, contains a vectorized facial image of size $19 \times 19$, and is not binary but satisfies $\mathbf{X}(i,j) \in [0,1]$ for all $(i,j)$. We binarize it by a simple rounding operation. 
Table~\ref{CBCL_all} reports the relative errors for $r=10$ and $r=20$. 
MS-Comb-AO, Tree BMF and the methods in \cite{Avellaneda_Villemaire_2022} are not reported in Table~\ref{CBCL_all} due to high memory and/or computational demands. We tried running the methods in \cite{Avellaneda_Villemaire_2022} for $r = 10$. Initially, we tested solely \texttt{FastUndercover}. Unfortunately, after 2 days, the method did not finish. As a result, we could not test the more demanding, in terms of time and memory, \texttt{optiblock}, neither could we test any of these methods for $r = 20$. MS-Comb-AO and Tree BMF require several AO-BMF solutions which is too slow (for one iteration, about 100 s.\ for $r=10$ and 500 s.\ for $r=20$). 
For MS-AO, we used 10 runs of AO-BMF with NMF-based initializations which was performing better.  


\begin{table}[h]
    \centering
    \begin{tabular}{lll} 
 \hline
 Method & \makecell{Best result\\ for $r = 10$} & \makecell{Best result\\ for $r = 20$}\\ [0.5ex] 
 \hline\hline
 MS-AO & \makecell{\textbf{59.61\%} \\ (1319s)} & \makecell{\textbf{54.74\%} \\ (4234s)}\\ 
 \hline
 Greedy-Comb & \makecell{60.20\% \\ (48s)} & \makecell{56.75\% \\ (222s)} \\
 \hline
 Greedy-TreeBMF & \makecell{\underline{60.18\%} \\ (56s)} & \makecell{\underline{56.65\%} \\ (160s)} \\
 \hline
 ELBMF \cite{dalleiger2022efficiently} & \makecell{74.66\% \\ (8s)} & \makecell{69.66\% \\ (19s)} \\
 \hline
 ASSO \cite{miettinen2008discrete} & \makecell{67.19\% \\ (23s)} & \makecell{65.62\% \\ (37s)} \\
 \hline
\end{tabular}
\caption{Relative errors for the binarized CBCL dataset. In parenthesis is the mean time over ten Monte Carlo trials. In bold is the best performing method, underlined is the second best.     \label{CBCL_all}} 
\end{table}
The AO-BMF method performs the best in terms of relative error but requires a very long runtime. For our greedy methods we considered the following parameters: 
\begin{itemize}
    \item Greedy-Comb: For $r= 10$, we gathered 5 solutions, while for $r = 20$ we gathered 15 solutions.
    \item Greedy-TreeBMF: For $r = 10$, we set a maximum timelimit of 50 seconds and solve 5 Greedy-Comb problems that each have a timelimit of 10 seconds. For $r = 20$, we set a maximum timelimit of 150 seconds and solve 5 Greedy-Comb problems that each have a timelimit of 30 seconds.
\end{itemize}
Our greedy-based methods are close, with Greedy-TreeBMF being slightly better than Greedy-Comb. 
%
The parameters used for ELBMF are the default ones. 
For ASSO, we set the threshold to 0.7. ELBMF and ASSO perform significantly worse than our proposed methods in terms of relative error, although they are faster, especially for $r=20$.

\begin{figure}[ht!]
	\begin{center} 
    \begin{tabular}{cc}
      \includegraphics[trim=0 30 0 0, clip, width=0.48\textwidth]{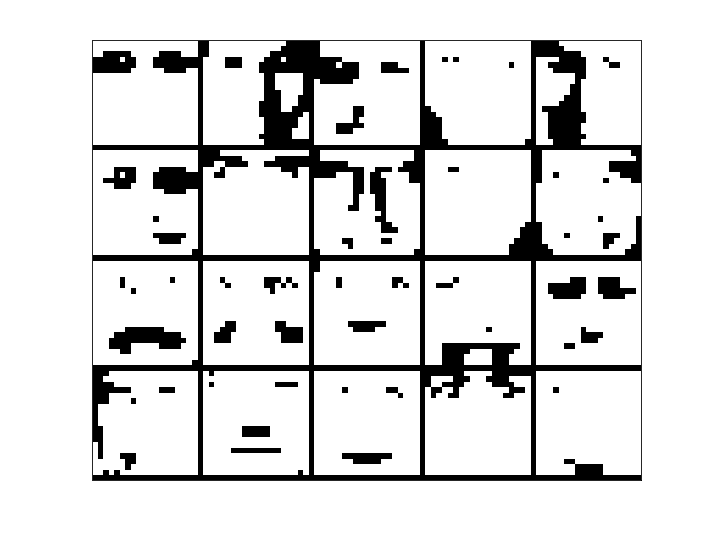}  & \includegraphics[trim=0 30 0 26, clip, width=0.48\textwidth]{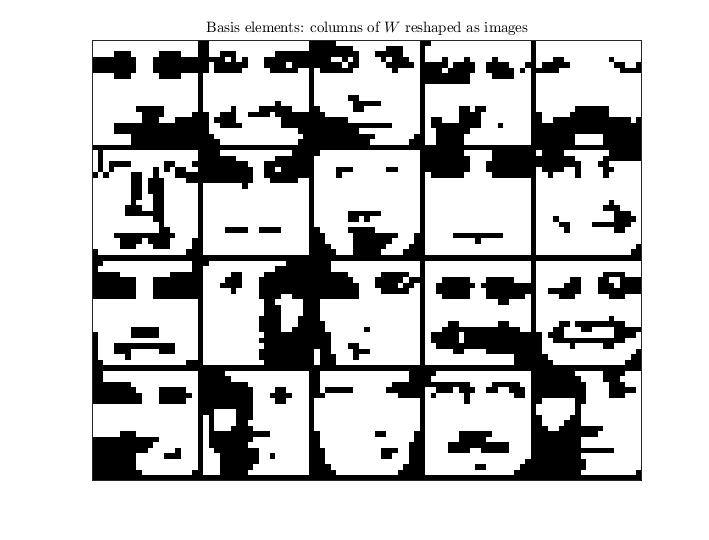} 
    \end{tabular}
		\caption{Facial features extracted by AO-BMF (left) and  Greedy-TreeBMF (right) on the binarized CBCL dataset. 
       \label{cbclim_greedy_hier}}   
	\end{center}
\end{figure}

 Figure~\ref{cbclim_greedy_hier} shows the facial features extracted by AO-BMF and Greedy-TreeBMF for $r=20$, which can easily be interpreted as common facial features from the dataset.  
 Interestingly, the facial features are rather different, although the relative errors are relatively close (54.74\% vs.\ 56.65\%). AO-BMF extracts more localized features while Greedy-TreeBMF extracts more global ones. The reason probably comes from the initialization procedure: AO-BMF uses NMF-based initialization that generates localized features, while Greedy-TreeBMF relies on a subset of the (dense) columns of $\bf X$ to initialize~$\bf W$. 

\section{Conclusion}

In this paper, we proposed novel algorithms to solve Boolean matrix factorization (BMF), with and without missing data. 
We first presented IP-based methods that solve BMF using alternating optimization over the factors, $\bf W$ and $\bf H$, along with strategies to combine several solutions. 
We then presented scalable, greedy-based methods that perform competitively while running significantly faster.  
Our extensive numerical experiments showed that our proposed algorithm outperforms the state of the art on small, medium, and large datasets. 

\newpage 



\small 

\bibliographystyle{spmpsci}
\bibliography{BMFsdp}



\normalsize 

\appendix 

\section{Modifying Greedy-Comb to enhance the diversity of solutions}  \label{app:enhanceprocedure}

 In our experiments, we noticed that several topics may be mined multiple times (that is, most words belonging to two columns of $\mathbf W$ are the same). 
For this reason, in order to obtain more diverse topics, we introduce an additional procedure, which is performed as an additional step at the end of  Greedy-Comb. 

After getting an initial solution $(\mathbf{W},\mathbf{H})$, the quantity $(\mathbf{H}^\top \mathbf{H})_{i,j}$ is a measure of how similar the topics $i$ and $j$ are, as it counts the number of common documents between these two topics. 
When $i = j$, this represents the number of documents that are associated with topic $i$. 
Ideally, we want the values on the diagonal to be large,  
while we would like the non-diagonal values to be as close to zero as possible, which corresponds to diverse topics.
To this end, we propose the following procedure: 
\begin{itemize}
    \item We first check the diagonal elements of $\mathbf{H}^\top \mathbf{H}$. If a diagonal element is less than the parameter $w$, 
   it removes it from the collection of solutions $\mathbf{U}$. We select a new solution through a modification of Algorithm~\ref{alg:rank1recon}, where instead of finding a completely new $\bf h$, we only substitute the index that has been removed. %
We used $w = 10$ in our experiments, which worked well for the large document dataset we considered.  
    
    \item If all diagonal elements are larger than or equal to $w$, we move to the next step. 
    If we find a non-diagonal element $(\mathbf{H}^\top \mathbf{H})_{i,j}$ close to $(\mathbf{H}^\top \mathbf{H})_{i,i}$, we also return the corresponding index and remove it from $\mathbf{U}$. This means that topics $i$ and $j$ contain many common documents, something that we want to avoid. Our measure of closeness is the ratio $\frac{(\mathbf{H}^\top \mathbf{H})_{i,i}}{(\mathbf{H}^\top \mathbf{H})_{i,j}}$, and a threshold of 8 worked well in our experiment. 
    
\end{itemize}
 This procedure continues until either: The conditions above are satisfied, or the number of remaining rank-one factors in the collection is equal to $r$, in which case we return the solution.

\end{document}